\documentclass[a4paper,twocolumn,english,prb,superscriptaddress]{revtex4}
\usepackage[T1]{fontenc}
\usepackage[latin9]{inputenc}
\usepackage{color}
\usepackage{bm}
\usepackage{amsmath}
\usepackage{amssymb}
\usepackage{graphicx}
\PassOptionsToPackage{normalem}{ulem}
\usepackage{ulem}

\makeatletter

\@ifundefined{textcolor}{}
{%
 \definecolor{BLACK}{gray}{0}
 \definecolor{WHITE}{gray}{1}
 \definecolor{RED}{rgb}{1,0,0}
 \definecolor{GREEN}{rgb}{0,1,0}
 \definecolor{BLUE}{rgb}{0,0,1}
 \definecolor{CYAN}{cmyk}{1,0,0,0}
 \definecolor{MAGENTA}{cmyk}{0,1,0,0}
 \definecolor{YELLOW}{cmyk}{0,0,1,0}
}

\makeatother

\usepackage{babel}
\begin{document}
\title{Waveguiding valley excitons in monolayer transition metal dicalcogenides by dielectric interfaces in the substrate}
\author{Xu-Chen Yang}
\affiliation{Department of Physics, The University of Hong Kong, Hong Kong, China}
\affiliation{HKU-UCAS Joint Institute of Theoretical and Computational Physics at Hong Kong, China}
\author{Hongyi Yu}
\email{yuhy33@mail.sysu.edu.cn}
\affiliation{Guangdong Provincial Key Laboratory of Quantum Metrology and Sensing and School of Physics and Astronomy, Sun Yat-Sen University (Zhuhai Campus), Zhuhai 519082, China}
\author{Wang Yao}
\affiliation{Department of Physics, The University of Hong Kong, Hong Kong, China}
\affiliation{HKU-UCAS Joint Institute of Theoretical and Computational Physics at Hong Kong, China}
\date{\today}
\begin{abstract}
In monolayers of the semiconducting transition metal dichalcogenides, the electron-hole exchange interaction splits the exciton dispersion into a massive transverse branch, and a longitudinal branch that has very light or even zero mass depending on the form of screened Coulomb interaction. The group velocity of the longitudinal branch is sensitive to the strength of electron-hole exchange, which can be engineered through the dielectric environment. Here we show that dielectric patterning of the substrate can be exploited to realize waveguide of the exciton in the longitudinal branch in a homogeneous monolayer, leaving the massive transverse branch unaffected. At a lateral interface of different dielectric constant in the substrate, the transmission and reflection of exciton in the longitudinal branch obey the Snell-Descartes law of optical system, and total reflection can be exploited to realize excitonic waveguide using two parallel interfaces. The same dielectric pattern of the substrate appears to be completely transparent for the massive transverse branch exciton, which has no interface scattering. When the monolayer is placed on a one-dimensional dielectric superlattice, the dispersion of the longitudinal branch is strongly renormalized, and the wavefunctions exhibit one-dimensional features, confined to either the low-dielectric or high-dielectric regions. In contrast, the massive transverse branch excitons are not affected by the substrate dielectric pattern, exhibiting pristine properties as in a freestanding monolayer.
\end{abstract}
\maketitle

\section{Introduction}

Atomically thin group\textendash VIB transition metal dichalcogenides (TMDs) has attracted intensive research interest in the recent decade \cite{Wang12,Xu14,Liu15}. The monolayer TMDs is a covalently bonded 2D hexagonal lattice, serving as the elementary unit of the layered bulk crystals stacked by the weak van der Waals interaction. In the atomically thin limit, the electronic band structure of the TMDs exhibits direct band gaps in visible frequency range\cite{Mak10,Splendiani10}, located at the energetically degenerate corners of the hexagonal Brillouin zone. So the conduction- and valence-band edges both have a degeneracy labeled by the valley index $\pm\textbf{K}$. An exciton is a hydrogen-like bound state of the band edge electron and hole by the Coulomb interaction. The reduced screening of Coulomb interaction in 2D geometry results in a large binding energy \cite{Jones13,Ross13,Mak13}. The exciton wavefunction, on the other hand, is still the Wannier type, with a small Bohr radius in the order of nm, well localised in the valleys in momentum space \cite{Qiu13}. The valley optical selection rule of the interband transition dictates that an exciton at $\textbf{K}$ (or $-\textbf{K}$) valley interconverts with a $\sigma+$ (or $\sigma-$) circularly polarized photon only \cite{Xiao12,Yao08}. Because of the small Bohr radius, the Coulomb exchange interaction between the electron and hole causes a sizable valley-orbit coupling, which splits the exciton energy-momentum dispersion into two branches of in-plane valley pseudospins, coupled to photons of linear polarization longitudinal and transverse to exciton momentum respectively \cite{Yu14,Yu15}. 
The transverse branch has parabolic dispersion with the regular exciton mass (sum of that of electron and hole). The longitudinal branch can be either massless or having very light mass depending on the form of screened Coulomb interaction, and the group velocity is sensitive to the strength of electron-hole Coulomb exchange which is inversely proportional to the square of exciton Bohr radius \cite{Yu14,Yu15,Qiu15,Wu15}. The latter has implied interesting possibilities to explore unusual transport properties of excitons, reminiscent of the massless Dirac electron in graphene \cite{Katsnelson06,Cheianov06,Cheianov07}

The Coulomb interaction in the monolayer geometry is notably sensitive to the surrounding environment. Placing monolayer TMDs on different dielectric substrate leads to modulation of the Coulomb interaction strength, which can result in a significant change both in the quasiparticle bandgap and in the exciton binding energy \cite{Ugeda14,Bradley15,Latini15,Stier16,Ryou16,Raja17,Cho18}. The dielectric-dependent bandgap renormalization has been exploited to realize lateral heterojunctions in continuous TMDs monolayer placed on substrate where the two adjacent segments have distinct dielectric constants \cite{Utama19}. 
For excitons, the effect from modulation of dielectric screening can be more subtle. The renormalization in the quasiparticle band gap and that in the exciton binding energy from the dielectric screening by substrate can cancel each other \cite{Qiu13,Cho18,Gao16}. The exciton dispersion edge, i.e. energy at zero momentum, can be measured as a resonance in the optical spectra. For the 1s exciton, experiments have found that this resonance has negligible change on different dielectric substrate \cite{Raja19,Stier16,Xu20,Lin14}. On the other hand, the dielectric screening of Coulomb interaction can affect the magnitude of the valley-orbit splitting in the exciton dispersion away from the dispersion edge. These finite momentum exciton states are optically dark, so the effect does not directly manifest in the optical spectral. Nevertheless, the exciton transport properties can be strongly modulated, especially for the longitudinal branch, where the group velocity sensitively responds to the strength of the electron-hole Coulomb exchange \cite{Yu14,Yu15,Qiu15,Wu15}. This suggests an interesting possibility to tailor exciton properties in monolayer TMDs through patterning dielectric substrate. 

In this work, we show that dielectric patterning of the substrate can be exploited to guide the exciton transport in the longitudinal branch in a homogeneous monolayer, while leaving the massive transverse branch unaffected. At a lateral interface where the two adjacent segments of substrate have distinct dielectric constants, the transmission and reflection of exciton in the longitudinal branch obey the Snell-Descartes law of optical system. The critical angle of the total internal reflection is dependent on the ratio of dielectric constants of the two substrate segments. Using two parallel interfaces, a waveguide can thus be realized for the longitudinal branch exciton. In contrast, such dielectric interfaces of the substrate appear completely transparent for excitons in the transverse branch, which has no interface scattering. 
When the monolayer is placed on a one-dimensional dielectric superlattice, the dispersion of the longitudinal branch exciton is changed significantly, displaying isotropic and anisotropic energy minima as well as saddle points at the center and edge of the superlattice Brillouin zone. The wavefunctions can exhibit one-dimensional (1D) features, confined to either the low-dielectric or high-dielectric regions. The massive transverse branch excitons are not affected by the substrate dielectric pattern, exhibiting the same two-dimensional (2D) parabolic dispersion as in a freestanding monolayer. This creates an interesting hybrid of 1D and 2D excitons in a common homogeneous monolayer.

The rest of this paper is organized as follows. In Sec.II, we introduce the model of monolayer TMDs excitons on periodic patterned dielectric superlattice. Then the mini-band dispersions of the excitons are calculated and analyzed in Sec.III. Sec.IV presents numerical results of the exciton transmission and reflection at a single lateral interface where the two adjacent segments of substrate have distinct dielectric constants, and put forward a waveguide design for the massless longitudinal branch exciton using two such interfaces. In Sec.V, we present the spatial features of wavefunctions showing waveguide confinement of excitons on dielectric superlattice. Then we discuss the effect of weak energy modulation by the dielectric superlattice in Sec.VI. We summary in Sec.VII.

\section{Model}

\begin{figure}
    \begin{centering}
        \includegraphics[scale=0.42]{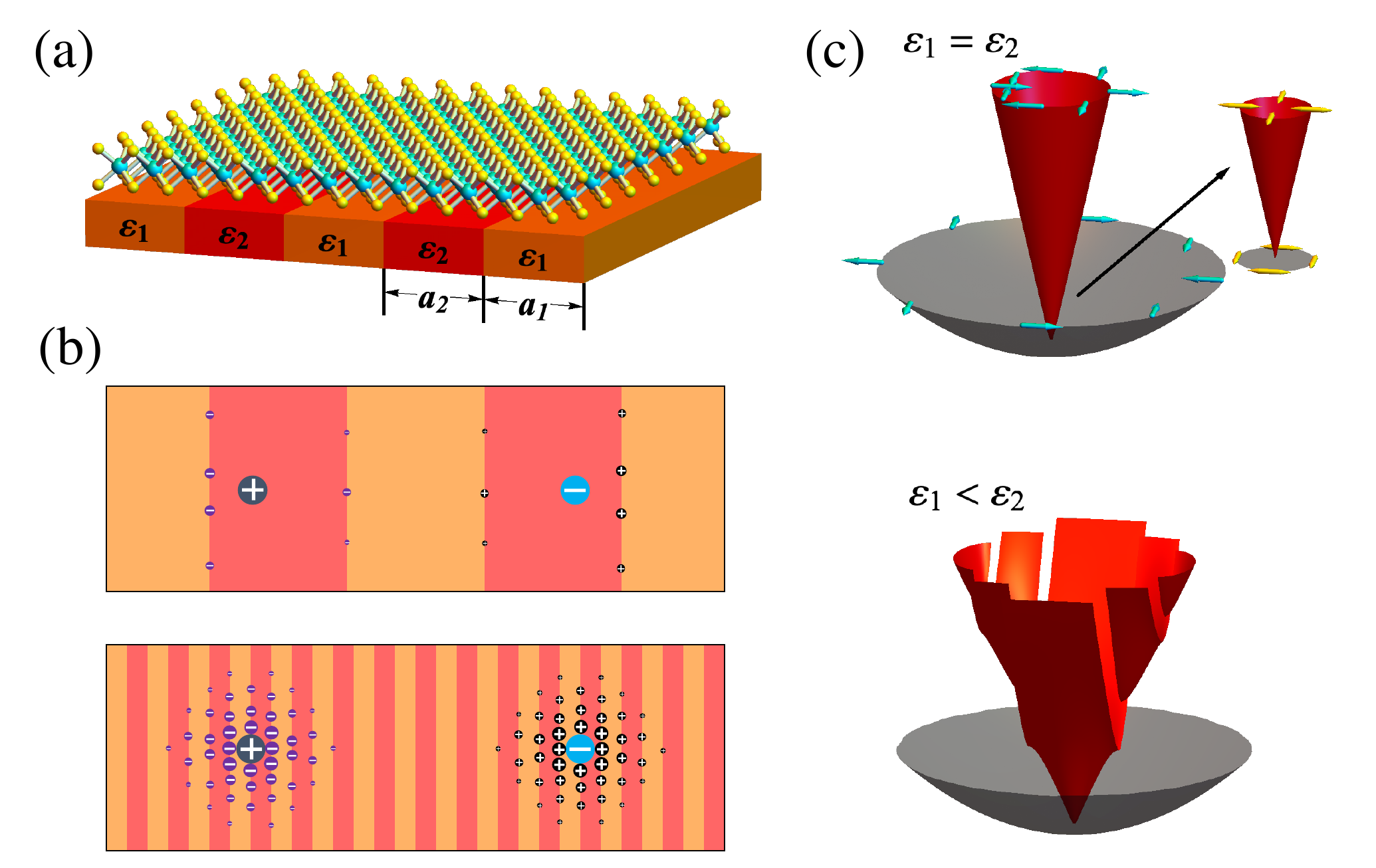}
        \par\end{centering}
        \caption{(a) Schematic diagrams of the monolayer TMDs on a substrate of a quasi-1D patterned dielectric superlattice $\varepsilon(\textbf{R})$, which is along the $x$-direction with period $L=a_1+a_2$. (b) The Coulomb interaction between the electron and hole is screened by the surface charges (cyan and pink dots) emerged at the dielectric interfaces, in the long and small period limits. (c) Energy dispersion of the exciton in homogeneous dielectric environment (top) and in dielectric superlattice $\varepsilon(\textbf{R})$ (bottom), where the red/gray color stands for the L/T branch. The cyan single-head arrows denote the valley pseudospin of excitons. In the inset of the dispersion in light cone, the yellow double-head arrows denote the linear polarization of the emitted photon on exciton recombination.
\label{fig:1despersion}}
\end{figure}

Here we focus on the spin-singlet bright excitons in monolayer TMDs with both the electron and hole constituents located at $+\textbf{K}$ or $-\textbf{K}$ valley. Meanwhile the spins of the electron and hole are opposite so the excitonic valley index is locked with its real spin ($+\textbf{K}/-\textbf{K}$ corresponding to $\uparrow/\downarrow$), due to the large spin-orbital splitting from the transition metal atom. The Coulomb exchange interaction between the electron and hole leads to the coupling between the two valleys, which breaks the valley degeneracy at finite center-of-mass (COM) momenta. The exchange splitting depends sensitively on the Coulomb interaction, which can be modulated spatially by placing the monolayer TMDs on a substrate with a patterned dielectric superlattice, schematically shown in Fig. \ref{fig:1despersion}(a). A general form of the Coulomb interaction can be written as $V(\textbf{R},\textbf{r}_\text{eh})=\frac{\text{e}^2}{\varepsilon(\textbf{R},\textbf{r}_\text{eh}r_\text{eh})}$, with $\varepsilon(\textbf{R},\textbf{r}_\text{eh})$ the effective dielectric screening experienced by the electron-hole pair, which depends on the electron-hole separation (relative coordinate) $\textbf{r}_\text{eh}\equiv\textbf{r}_\text{e}-\textbf{r}_\text{h}$ \cite{Chernikov14}, and now also spatially varies with the COM coordinate $\textbf{R}\equiv\frac{m_\text{e}}{M_0}\textbf{r}_\text{e}+\frac{m_\text{h}}{M_0}\textbf{r}_\text{h}$ due to the patterned dielectric. $\textbf{r}_{\text{e}/\text{h}}$ denotes the electron/hole spatial coordinate.

In the valley pseudospin space, the exchange interaction can be written as
\begin{equation}
    \hat{H}_\text{ex}=\sum_{\textbf{k},\textbf{k}^\prime,\tau,\tau^\prime}J_{\tau\tau^\prime}\left(\textbf{k},\textbf{k}^\prime\right)\hat{B}_{\textbf{k},\tau}^\dagger\hat{B}_{\textbf{k}^\prime,\tau^\prime}^{}
    \label{eq:totH}
\end{equation}
Here $\hat{B}_{\textbf{k},\tau}^\dagger$ ($\hat{B}_{\textbf{k},\tau}^{}$) creates (annihilates) an exciton with the COM momentum $\textbf{k}=\left(k_x,k_y\right)=\left(k\cos\theta,\sin\theta\right)$ and the valley pseudospin $\tau=+,-$, which can be described by the wavefunction $X_{\textbf{k},\tau}(\textbf{r}_\text{e},\textbf{r}_\text{h})\equiv\sum_\textbf{q}\Phi(\textbf{q})\psi_{\textbf{q}+\frac{m_\text{e}}{M_0}\textbf{k},\text{c},\tau}(\text{r}_\text{e})^{}\psi_{\textbf{q}-\frac{m_\text{h}}{M_0}\textbf{k},\text{v},\tau}^*(\text{r}_\text{h})$. $\psi_{\textbf{k},\text{c}/\text{v},\tau}(\textbf{r}_{\text{e}/\text{h}})$ is the conduction/valence band Bloch state at a wave vector away from $\tau\textbf{K}$ by $\textbf{k}$. The exciton mass $M_0=m_\text{e}+m_\text{h}\approx0.7m_0$ with $m_\text{e}\approx m_\text{h}$ and $m_0$ is the free electron mass. $J_{+-/-+}(J_{++/--})$ is the strength of the intervalley (intravalley) Coulomb exchange interaction whose ampitude is (see Appendix~\ref{appxA} for detail)
\begin{equation}
    \begin{aligned}
        &J_{\tau,\tau^\prime}(\textbf{k},\textbf{k}^\prime)\\
        =&\int\text{d}\textbf{r}_\text{e}\text{d}\textbf{r}_\text{h}X^*_{\textbf{k},\tau}(\textbf{r}_\text{e},\textbf{r}_\text{e})V(\textbf{R},\textbf{r}_\text{eh})X^{}_{\textbf{k}^\prime,\tau^\prime}(\textbf{r}_\text{h},\textbf{r}_\text{h})\\
        =&\left(\textbf{k}\cdot\textbf{d}_{X,\tau}\right)\left(\textbf{k}^\prime\cdot\textbf{d}^*_{X,\tau^\prime}\right)V\left(\textbf{k}-\textbf{k}^\prime,\frac{m_\text{e}}{M_0}\textbf{k}+\frac{m_\text{e}}{M_0}\textbf{k}^\prime\right)
    \end{aligned}
    \label{eq:exchange}
\end{equation}
In the above we have dropped the short-ranged part of the exchange interaction, as it depends weakly on the momenta thus can be views as a constant for small values of $k$ and $k^\prime$ (see Appendix~\ref{appxA} for detail). $\textbf{d}_{X,\tau}$ is the optical dipole of the exciton, and
\begin{equation}
    V(\textbf{Q},\textbf{k})\equiv\int\text{d}\textbf{r}_\text{eh}\text{d}\textbf{R}\text{e}^{-\text{i}\textbf{k}\cdot\textbf{r}_\text{eh}}\text{e}^{-\text{i}\textbf{Q}\cdot\textbf{R}}V(\textbf{R},\textbf{r}_\text{eh})
    \label{eq:V}
\end{equation}
As the Coulomb interaction $V(\textbf{R},\textbf{r}_\text{eh})$ now varies with the COM coordinate $\textbf{R}$, generally $J_{\tau,\tau^\prime}(\textbf{k},\textbf{k}^\prime)\neq0$ for $\textbf{k}\neq\textbf{k}^\prime$.

In monolayer TMDs, the valley optical selection rule indicates that the exciton in $+\textbf{K}$ (or $-\textbf{K}$) valley can only interconvert with a $\sigma+$ (or $\sigma-$) circularly polarized photon. Such a circularly polarized selection rule originates from the $2\pi/3$-rotational symmetry of the lattice, and is expected to hold as long as the length scale of spatial modulation of the dielectric constant is much larger than the monolayer lattice constant. So the optical dipole of the $\tau\textbf{K}$ valley exciton is $\textbf{d}_{X,\tau}\equiv\left(d^x_{X,\tau},d^y_{X,\tau}\right)=\frac{D}{\sqrt{2}}(\text{i}\tau,1)$ \cite{Yu14}, and 
\begin{equation}
    \begin{aligned}
        &J_{\tau,\tau^\prime}(\textbf{k},\textbf{k}^\prime)\\
        =&\frac{D^2}{2}kk^\prime\left(\text{e}^{-\text{i}\tau\theta+\text{i}\tau^\prime\theta^\prime}\right)V\left(\textbf{k}-\textbf{k}^\prime,\frac{m_\text{e}}{M_0}\textbf{k}+\frac{m_\text{e}}{M_0}\textbf{k}^\prime\right)
    \end{aligned}
    \label{eq:exchange1}
\end{equation}
The exchange Hamiltonian then becomes
\begin{equation}
        \hat{H}_\text{ex}=D^2\sum_{\textbf{k},\textbf{k}^\prime}kk^\prime V\left(\textbf{k}-\textbf{k}^\prime,\frac{m_\text{e}}{M_0}\textbf{k}+\frac{m_\text{e}}{M_0}\textbf{k}^\prime\right)\hat{B}^\dagger_{\textbf{k},\text{L}}\hat{B}^{}_{\textbf{k}^\prime,\text{L}}
    \label{eq:Hex}
\end{equation}
Here $\hat{B}^\dagger_{\textbf{k},\text{L}}\equiv\frac{1}{\sqrt{2}}\left(\text{e}^{-\text{i}\theta_\textbf{k}}\hat{B}^\dagger_{\textbf{k},+}+\text{e}^{\text{i}\theta_\textbf{k}}\hat{B}^\dagger_{\textbf{k},-}\right)$ and $\hat{B}^\dagger_{\textbf{k},\text{T}}\equiv\frac{1}{\sqrt{2}}\left(\text{e}^{-\text{i}\theta_\textbf{k}}\hat{B}^\dagger_{\textbf{k},+}-\text{e}^{\text{i}\theta_\textbf{k}}\hat{B}^\dagger_{\textbf{k},-}\right)$ are the creation operators of the longitudinal and transverse branches, respectively, which emit linearly polarized photons with polarization angles longitudinal and transverse to the direction of $\textbf{k}$.

Interestingly, due to the circularly polarized valley optical selection rule the Coulomb exchange interaction only affects the longitudinal branch, while the T branch is totally unaffected. Note that this conclusion is independent on the form of the Coulomb interaction $V\left(\textbf{R},\textbf{r}_\text{eh}\right)$.

Below, we consider one-dimensional modulation of the dielectric, i.e. the local dielectric constant $\varepsilon(\textbf{R})\equiv\varepsilon\left(\textbf{R},\textbf{r}_\text{eh}\rightarrow0\right)$ periodically varies in the $x$-direction only. In a period $x\in\left[-L/2,L/2\right]$,
\begin{equation}
\varepsilon\left(\textbf{R}\right)=\begin{cases}
\begin{array}{ccc}
\varepsilon_{1}=\varepsilon_{0}, &  & -a_{1}/2\leq x<a_{1}/2,\\
\varepsilon_{2}=\xi\cdot\varepsilon_{1}, &  & \text{otherwise},
\end{array}\end{cases}\label{eq:1Dep}
\end{equation}
which is schematically shown in Fig.\ref{fig:1despersion} (a). We emphasize that for an electron-hole pair separated by  $\textbf{r}_\text{eh}\neq0$, the effective dielectric screening $\varepsilon\left(\textbf{R},\textbf{r}_\text{eh}\right)$ of their Coulomb interaction can be different from the local value $\varepsilon(\textbf{R})$. Nevertheless, the dielectric constant $\varepsilon\left(\textbf{R},\textbf{r}_\text{eh}\right)$ always has the same periodicity as a function of the COM coordinate $\textbf{R}$, so $V\left(\textbf{Q},\textbf{k}\right)=\sum_\textbf{g}\delta_{\textbf{Q},\textbf{g}}V\left(\textbf{g},\textbf{k}\right)$, where $\textbf{g}$ is the reciprocal lattice vector of the dielectric superlattice.

\section{RENORMALIZATION OF THE EXCITON DISPERSION BY THE PATTERNED DIELECTRIC SUPERLATTICE}

The energy dispersion of the monolayer exciton is determined by the total Hamiltonian 
\begin{equation}
	\begin{aligned}
		\hat{H}=&\hat{H}_\textbf{k}+\hat{H}_\text{ex}\\
		=&\sum_\textbf{k}\frac{\hbar^2k^2}{2M_0}\left(\hat{B}^\dagger_{\textbf{k},\text{L}}\hat{B}^{}_{\textbf{k},\text{L}}+\hat{B}^\dagger_{\textbf{k},\text{T}}\hat{B}^{}_{\textbf{k},\text{T}}\right)\\
		&+D^2\sum_{\textbf{k},\textbf{k}^\prime}kk^\prime V\left(\textbf{k}-\textbf{k}^\prime,\frac{m_\text{e}}{M_0}\textbf{k}+\frac{m_\text{e}}{M_0}\textbf{k}^\prime\right)\hat{B}^\dagger_{\textbf{k},\text{L}}\hat{B}^{}_{\textbf{k}^\prime,\text{L}}.
	\end{aligned}
\end{equation}
The Hamiltonian indicates that the dispersion of the L branch is modulated by the patterned dielectric substrate, while the dispersion of the L branch keeps unchanged.

In the case of homogeneous dielectric substrate $\varepsilon_2=\varepsilon_1$ (i.e., $\xi=1$), the Coulomb potential $V(\textbf{R},\textbf{r}_\text{eh})=V(\textbf{r}_\text{eh})$ depends on $\textbf{r}_\text{eh}$ only, and $V(\textbf{Q},\textbf{k})=\delta_{\textbf{Q},0}V(\textbf{k})$ with $V(\textbf{k})\equiv\int\text{d}\textbf{r}_\text{eh}\text{e}^{-\text{i}\textbf{k}\cdot\textbf{r}_\text{eh}}V(\textbf{r}_\text{eh})=\frac{2\pi\text{e}^2}{\varepsilon_0(k+k_\text{s})}$ when taking into account the screening effect from free carriers. The resultant exchange interaction only couples two excitons with the same COM momentum,
\begin{equation}
    \hat{H}_\text{ex}=\frac{2\pi\text{e}^2D^2}{\varepsilon_0}\sum_\textbf{k}\frac{k^2}{k+k_\text{s}}\hat{B}^\dagger_{\textbf{k},\text{L}}\hat{B}^{}_{\textbf{k},\text{L}}.
\label{eq:exH}
\end{equation}

The result illustrates that the exciton's energy band is split into two branches of in-plane valley pseudospins, coupled to photons of linear polarization longitudinal and transverse branch to the exciton momentum respectively \cite{Yu14,Yu15}. The T branch has a parabolic dispersion with the regular exciton mass, while the L branch can be either massless or having a very light mass depending on the screening wavevector $k_\text{s}$. In the limit of low free carrier density, the screening wavevector $k_\text{s}=\frac{2m^*e^2}{\varepsilon_0\hbar^2}f_0$, where $f_0$ is the Fermi-Dirac distribution of the electron or hole at the $\pm\textbf{K}$ point. For $k_\text{s}=0$ the L branch has a linear dispersion near $\textbf{k}=0$. The dispersion of the exciton is shown in the upper panel of Fig.\ref{fig:1despersion} (c). 
\begin{figure}
\begin{centering}
\includegraphics[scale=0.36]{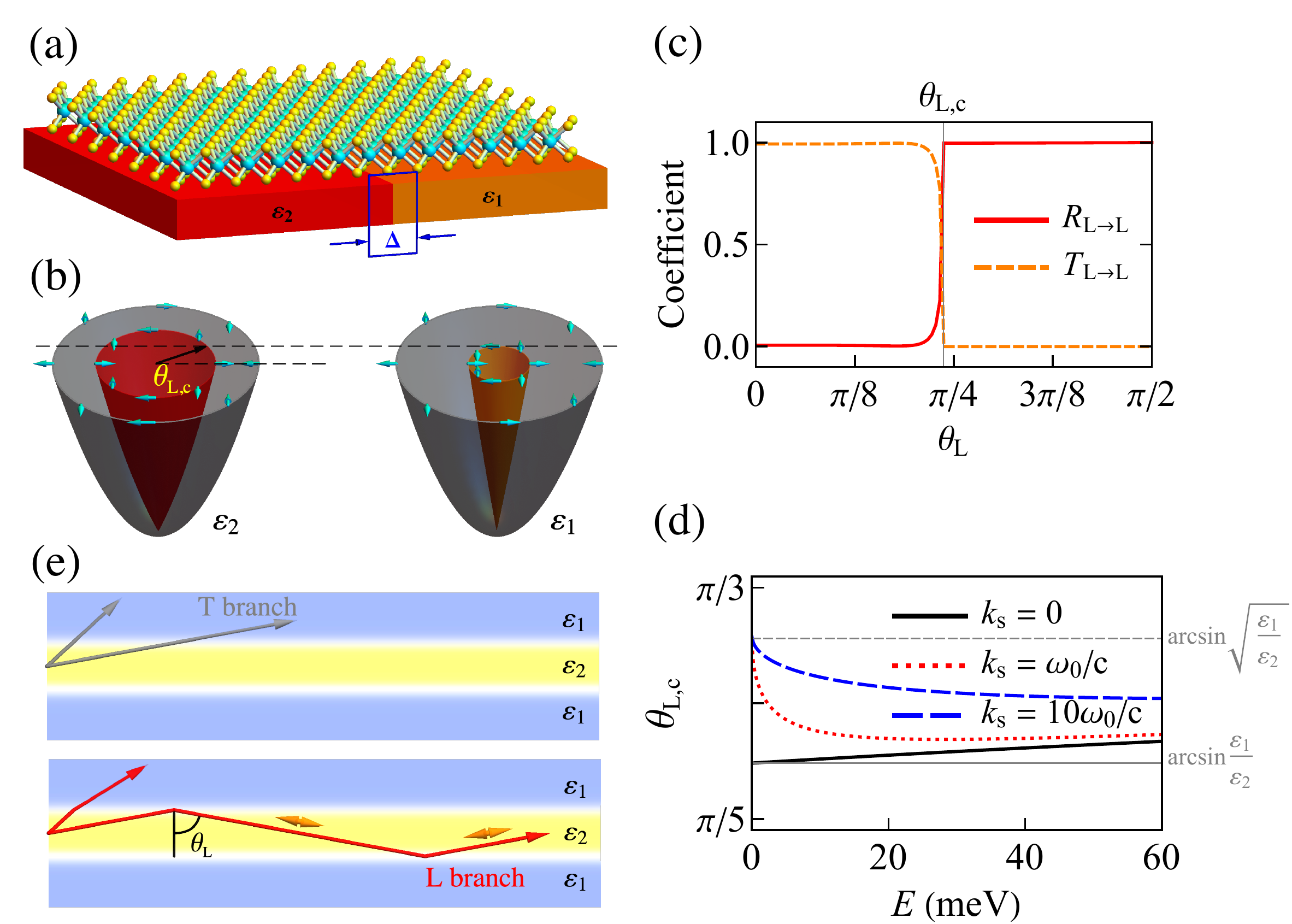}
\par\end{centering}
\caption{(a) Schematic diagrams of the monolayer TMDs at a lateral interface, where the two adjacent regions have distinct dielectric constants ($\varepsilon_1$ and $\varepsilon_2$). A transitional zone with widths $\Delta$ is at the interface. (b) Energy dispersion and valley pseudospin of the excitons at each region. The critical angle $\theta_\text{L,c}$ of the total internal reflection is marked. (c) Transmission and reflection of the L branch incidence as a function of incident angle $\theta_\text{L}$ in $k_\text{s}=0$. (d) $\theta_\text{L,c}$ as a function of the exciton energy $E$. For the black solid, red dotted and blue dashed lines, the screening wavevector $k_\text{s}=$0, 1, and 10, respectively, in units of $\omega_0/c$. The limit value of $\theta_\text{L,c}$, dashed gray line for $\arcsin(\varepsilon_1/\varepsilon_2)$ and solid gray line for $\arcsin\sqrt{\varepsilon_1/\varepsilon_2}$. (e) Schematic diagrams of an excitonic waveguide formed by three dielectric regions ($\varepsilon_1-\varepsilon_2-\varepsilon_1$). The propagating directions of L/T branch are denoted by red/gray single-head arrows. The yellow double-head arrows denote the linear polarization of the coupled photons. Parameters: $\Delta=20$ nm, $J=1$ eV, and $\xi_\varepsilon=1.5$.
\label{fig:2junction}}
\end{figure}

For an inhomogeneous dielectric superlattice with $\varepsilon_2\neq\varepsilon_1$, we can expand the effective dielectric constant into the Fourier series $\varepsilon^{-1}(\textbf{R},\textbf{r}_\text{eh})=\varepsilon^{-1}_0\sum_\textbf{g}\alpha(\textbf{g},\textbf{r}_\text{eh})\text{e}^{\text{i}\textbf{g}\cdot\textbf{R}}$. In this case $V(\textbf{g},\textbf{k})=\frac{\text{e}^2}{\varepsilon_0}\int\text{d}\textbf{r}_\text{eh}\alpha(\textbf{g},\textbf{r}_\text{eh})\frac{\text{e}^{-\text{i}\textbf{k}\cdot\textbf{r}_\text{eh}}}{r_\text{eh}}$ is generally nonzero for $\textbf{g}\neq0$. Note that surface charges can appear at the dielectric interface between $\varepsilon_1$ and $\varepsilon_2$, which can effectively screen the Coulomb interaction between the electron and hole, as schematically shown in Fig.\ref{fig:1despersion} (b). Such a mechanism is analogous to the effect from the free carriers, thus we write the Coulomb interaction form as $V(\textbf{g},\textbf{k})=\alpha(\textbf{g})\frac{2\pi\text{e}^2}{\varepsilon_0}\frac{1}{k_0(\textbf{g})+k}$ with $\alpha(\textbf{g})\equiv\alpha(\textbf{g},\textbf{r}_\text{eh}=0)$. $k_0(\textbf{g})$ is the screening wavevector from the surface charges, which is a function of the modulation period $L$ or the reciprocal lattice vector $\textbf{g}$. In the long period limit ($L\rightarrow\infty$) the Coulomb interaction becomes the unscreened form, i.e., $k_0(\textbf{g}=0)=0$. Furthermore the surface charges only appear at the dielectric interface, thus there are more charges for a smaller $L$ where the screening becomes more efficient, implying that $k_0(\textbf{g})$ increases with $g$. Accordingly we adopt the form $k_0(\textbf{g})=\lambda g$ with $\lambda$ a tunable parameter, then the Coulomb exchange interaction is
\begin{equation}
    \hat{H}_\text{ex}=\frac{J}{K}\sum_{\textbf{k},\textbf{k}^\prime}\delta_{\textbf{k}^\prime,\textbf{k}+\textbf{g}}\alpha(\textbf{g})\frac{kk^\prime}{\lambda g+\left|\frac{m_\text{h}}{M_0}\textbf{k}+\frac{m\text{e}}{M_0}\textbf{k}^\prime\right|}\hat{B}^\dagger_{\textbf{k},\text{L}}\hat{B}^{}_{\textbf{k}^\prime,\text{L}}
\label{eq:exH1}
\end{equation}
with $J\equiv\frac{2\pi\text{e}^2D^2}{\varepsilon_0}K\sim1$ eV (see Appendix~\ref{appxA} for detail). The dispersion of the exciton is shown in the lower panel of Fig.\ref{fig:1despersion} (c).

\section{Scattering analysis at a lateral interface}
Here we first analyze the scattering of the exciton in monolayer at a lateral interface, where the two adjacent regions of the substrate (labeled by $\eta=1,2$) have distinct dielectric constants $\varepsilon_\eta$ (Fig.\ref{fig:2junction} (a)). At region $\eta$, the Hamiltonian of the exciton can be written as \cite{Yu15},
\begin{equation}
	\hat{H}_\eta=\frac{\hbar^2k^2}{2M_0}+J^\text{intra}_{\textbf{k},\eta}+\left[J^\text{inter}_{\textbf{k},\eta}\hat{\sigma}_++\text{h.c.}\right]
\label{eq:etaH}
\end{equation}
where the valley pseudospin raising/lower operators $\hat{\sigma}_\pm=\left(\hat{\sigma}_x\pm\text{i}\hat{\sigma}_y\right)/2$. The intervalley coupling is $J^\text{inter}_{\textbf{k},\eta}=\frac{\pi\text{e}^2D^2}{\varepsilon_\eta(k+k_\text{s})}k^2\text{e}^{-2\text{i}\theta}$ and intravalley coupling is $J^\text{intra}_{\textbf{k},\eta}=\left|J^\text{inter}_{\textbf{k},\eta}\right|$. $\hat{H}_\eta$ splits the exciton into the L- and T-branches. The eigenstates can be written as the direct product of the orbital and valley pseudospin parts $\left|\psi_{\eta,\chi}(\textbf{r})\right\rangle=\left|\phi_{\eta}(\textbf{r})\right\rangle\otimes\left|\chi\right\rangle$, $\chi=\text{L}, \text{T}$. Fig.\ref{fig:2junction} (b) shows the energy dispersion and valley pseudospin of the exciton at the two regions.

While the interface can be sharp in the substrate, there can be a finite length scale for the variation of the local dielectric constant experienced by the exciton in the monolayer, given the finite vertical distance and the exciton radius. Therefore, we considered a $\Delta$-width transitional region where the local dielectric constant changes continuously from the value of the left region to that value of the right region. In the scattering analysis, this transitional region is divided into $n$ equal pieces with the local dielectric constant $\left[i\varepsilon_1+(n-i)\varepsilon_2\right]/n$ at the $i$th piece, and the wavefunctions are matched with the boundary conditions (Eq.(\ref{eq:tmdbc1}) in Appendix~\ref{appxC}). The exciton state at each region can be written as $\left|\Psi_\eta(\textbf{r})\right\rangle=\sum_{\chi=\text{L},\text{T}}c_{\eta,\chi}\left|\psi_{\eta,\chi}(\textbf{r})\right\rangle$, where the coefficients $c_{\eta,\chi}$ are determined by boundary conditions. (see Appendix~\ref{appxC} for detail). With the translational invariance in the $y$ direction, $k_y$ is a good quantum number. The number of steps $n$ is increased until the results saturate, such that the scattering by a smooth transition of dielectric constant over length scale $\Delta$ is obtained. The $x$-component current density of $\chi$- branch exciton is  $j^x_{\eta,\chi}=\left|c_{\eta,\chi}\right|^2\left\langle\psi_{\eta,\chi}(\textbf{r})\right|\hat{v}_{\eta,x}\left|\psi_{\eta,\chi}(\textbf{r})\right\rangle$, where the velocity operator is $\hat{v}_{\eta,x}=\left[x,\hat{H}_\eta\right]/\text{i}\hbar$ \cite{Matsuyama02}. The coefficient $R_{\chi\rightarrow\chi^\prime}$ ($T_{\chi\rightarrow\chi^\prime}$) is defined as the ratio between the reflection (transmission) current of the $\chi^\prime$ branch and the incident current of $\chi$ branch.

\begin{figure}
\begin{centering}
\includegraphics[scale=0.3]{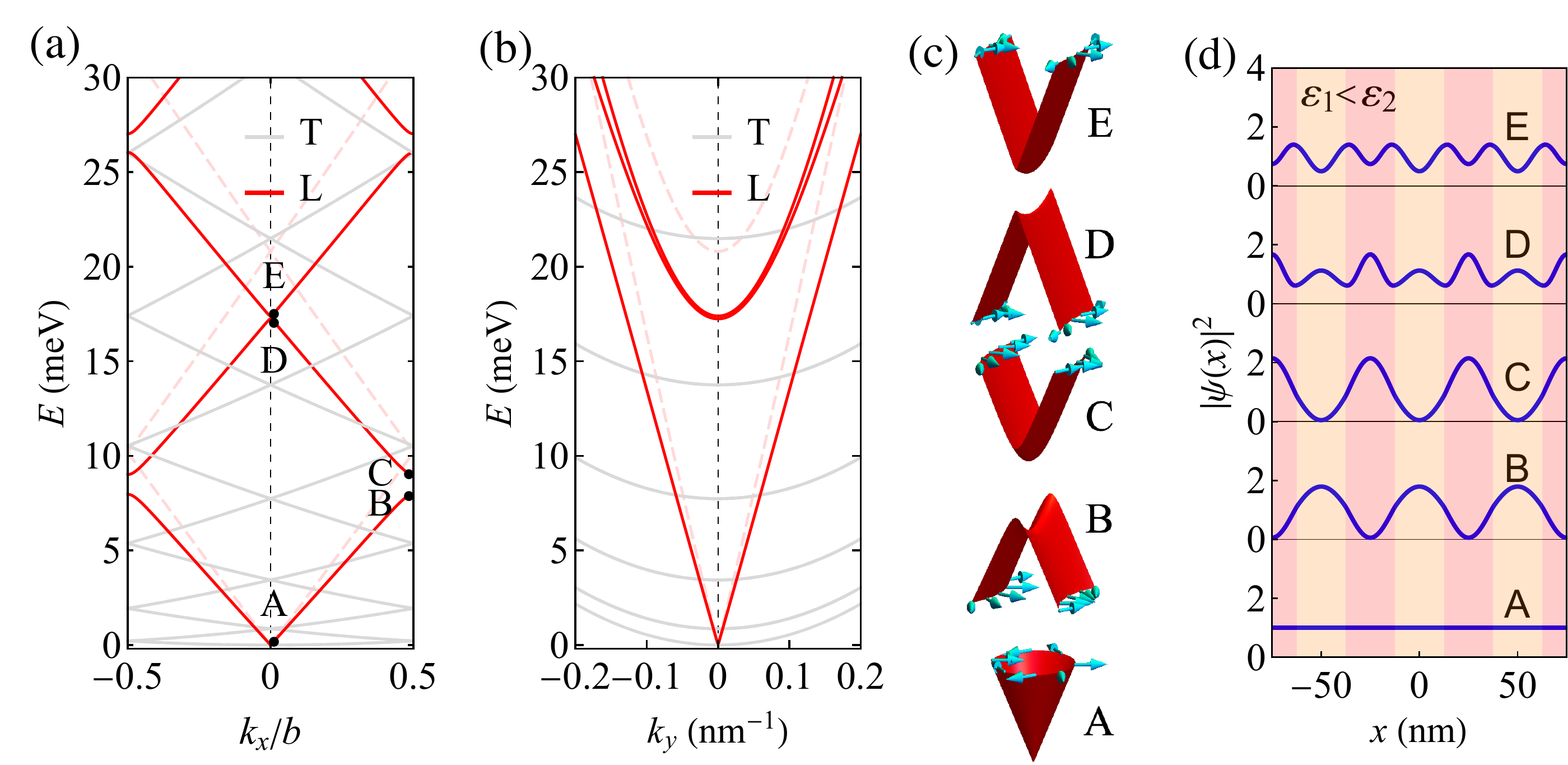}
\par\end{centering}
\caption{Renormalized energy dispersion of the exciton in 1D dielectric superlattice $\varepsilon(\textbf{R})$, where the red/gray solid lines stand for the L/T branch. Dispersion in the homogeneous dielectric environment is denoted by the dashed lines. (a) Dispersion as functions of $k_x$ at $k_y=0$. (b) Dispersion as functions of $k_y$ at $k_x=0$. (c) Dispersions and valley pseudospin of the L branch exciton near the centers (A, D, E) and edges (B, C) of the superlattice Brillouin zone. (d) 1D features of the L branch exciton's wavefunction. Parameters: $J=1$ eV, $\xi_\varepsilon=1.5$, and $a_1=a_2=25$ nm.
\label{fig:3wavefunction}}
\end{figure}

Our calculation finds that the T-branch exciton is not affected by the dielectric lateral interface, i.e., $T_{\text{T}\rightarrow\text{T}}=1$ and $R_{\text{T}\rightarrow\text{T}}=0$, consistent with the fact that the T-branch dispersion is unchanged across the dielectric interface. Moreover, scattering between the T- and L-branches is insignificant. The L-branch is significantly affected. $T_{\text{L}\rightarrow\text{L}}$ and $R_{\text{L}\rightarrow\text{L}}$ are shown in Fig.\ref{fig:2junction} (c) as functions of the incident angle $\theta_\text{L}$ in the unscreened limit ($k_\text{s}=0$). $T_{\text{L}\rightarrow\text{L}}$ is close to unit for a large range of incident angle around the normal incidence. When $\theta_\text{L}$ is above a critical angle $\theta_\text{L,c}$, $T_{\text{L}\rightarrow\text{L}}$ suddenly drops to zero, while $R_{\text{L}\rightarrow\text{L}}$ rises quickly to unit, becoming the dominant outgoing channel. This phenomenon is similar to the phenomenon of light passing through an interface with different media in optical system. The incident angle $\theta_\text{L}$ and the refraction angle obey the Snell-Descartes law. The critical angle $\theta_\text{L,c}$ depends on $k_\text{s}$ and exciton scattering energy $E$, as shown in Fig. \ref{fig:2junction} (d). In low energy limit, $\theta_\text{L,c}\approx\arcsin(\varepsilon_1/\varepsilon_2)$ at $k_\text{s}=0$, while $\theta_\text{L,c}\approx\arcsin\sqrt{\varepsilon_1/\varepsilon_2}$ at $k_\text{s}=10\omega_0/c$. The parameters adopted in the calculation are: $\Delta=20$ nm, $J=1$ eV, $\hbar\omega_0=2$ meV, and $\xi_\varepsilon=1.5$.

The photon-like scattering behavior suggests that an excitonic waveguide can be realized for the L-branch excitons by patterning the substrate dielectric as $\varepsilon_1-\varepsilon_2-\varepsilon_1$ ($\varepsilon_1<\varepsilon_2$) with parallel two interfaces, as schematically shown in Fig. \ref{fig:2junction} (d). We consider an exciton wavepacket in region $\varepsilon_2$, whose angle of the propagating direction is $\theta_\text{L}$. While the T-branch exciton is always transmitted through the interface, the transmission or reflection of the L-branch is determined by the angle $\theta_\text{L}$. When $\theta_\text{L}>\theta_\text{L,c}$, the L-branch exciton has total reflection at the interface. Otherwise ($\theta_\text{L}<\theta_\text{L,c}$), it refracts outside. Thus the L-branch with small $\theta_\text{L}$ is confined in the dielectric region $\varepsilon_2$ and an excitonic waveguide is achieved.

\section{WAVEGUIDE CONFINEMENT OF EXCITON'S WAVEFUNCTION}
The scattering analysis in the previous section suggests waveguide confinement of exciton by two adjacent dielectric interfaces. Here we further examine this 1D propagation of the L-branch excitons, by solving the exciton wavefunction in a 1D dielectric superlattice consisting of periodically arranged parallel interfaces, where the dielectric coefficent $\varepsilon(\textbf{R})$ is described by Eq.(\ref{eq:1Dep}), as shown in Fig.\ref{fig:1despersion} (a) with $a_1=a_2=25$ nm, $\varepsilon_2=1.5\varepsilon_1$ and $\lambda=1$.

The renormalized energy dispersion of the exciton along $x$-axis ($k_y=0$) is shown in Fig.\ref{fig:3wavefunction} (a). While the T branch (gray solid lines) is not affected by the substrate dielectric pattern, the dispersion of the L branch exciton (red solid line) is changed significantly. For the lowest band near $\textbf{k}=0$, the L branch remains linear but has a renormalized group velocity compared to the homogeneous dielectric case (red dashed lines), and the reduced group velocity is proportional to $\xi_\varepsilon$. Furthermore, bandgap opens for the L branch at the center/edge of the superlattice Brillouin zone. Fig.\ref{fig:3wavefunction} (b) shows the dispersion along $y$-axis and $k_x=0$. Fig.\ref{fig:3wavefunction} (c) is an enlarged view for dispersions of the exciton L branch near the center (A, D, E) and edges (B, C) of the superlattice Brillouin zone. The L branch has isotropic energy minima and isotropic group velocity near A, but is anisotropic near C or B. For instance, saddle shapes of energy minima appear at B and D points.

The patterned dielectric superlattice also introduces one-dimensional (1D) modulations to the exciton's wavefunction. Fig.\ref{fig:3wavefunction} (a) shows the density distributions of the L branch at momentum-space positions A to E. At the lowest energy A point, the exciton density keeps constant in real-space, due to the fact that the exchange matrix element $J_{\tau,\tau^\prime}(\textbf{k},\textbf{k}^\prime)=0$ when $kk^\prime=0$. However the densities at other higher-energy positions exhibit spatial modulations, confined to either the low-dielectric or high-dielectric regions. For instance, the calculated density distributions indicate that the low-energy mode at B has dominant probability in the low-$\varepsilon$ regions, while the high-energy mode at C concentrates in the high-$\varepsilon$ regions. A similar phenomenon is presented between the 3rd band (D point) and the 4th band (E point) of the L exciton branch at the center of the superlattice Brillouin zone $k=0$. These spatial features are not sensitive to the smooth interface length $\Delta$, the screening wavevector parameter $\lambda$.

\section{EFFECTS OF ENERGY MODULATION IN THE DIELECTRIC SUPERLATTICE}

\begin{figure}
\begin{centering}
\includegraphics[scale=0.39]{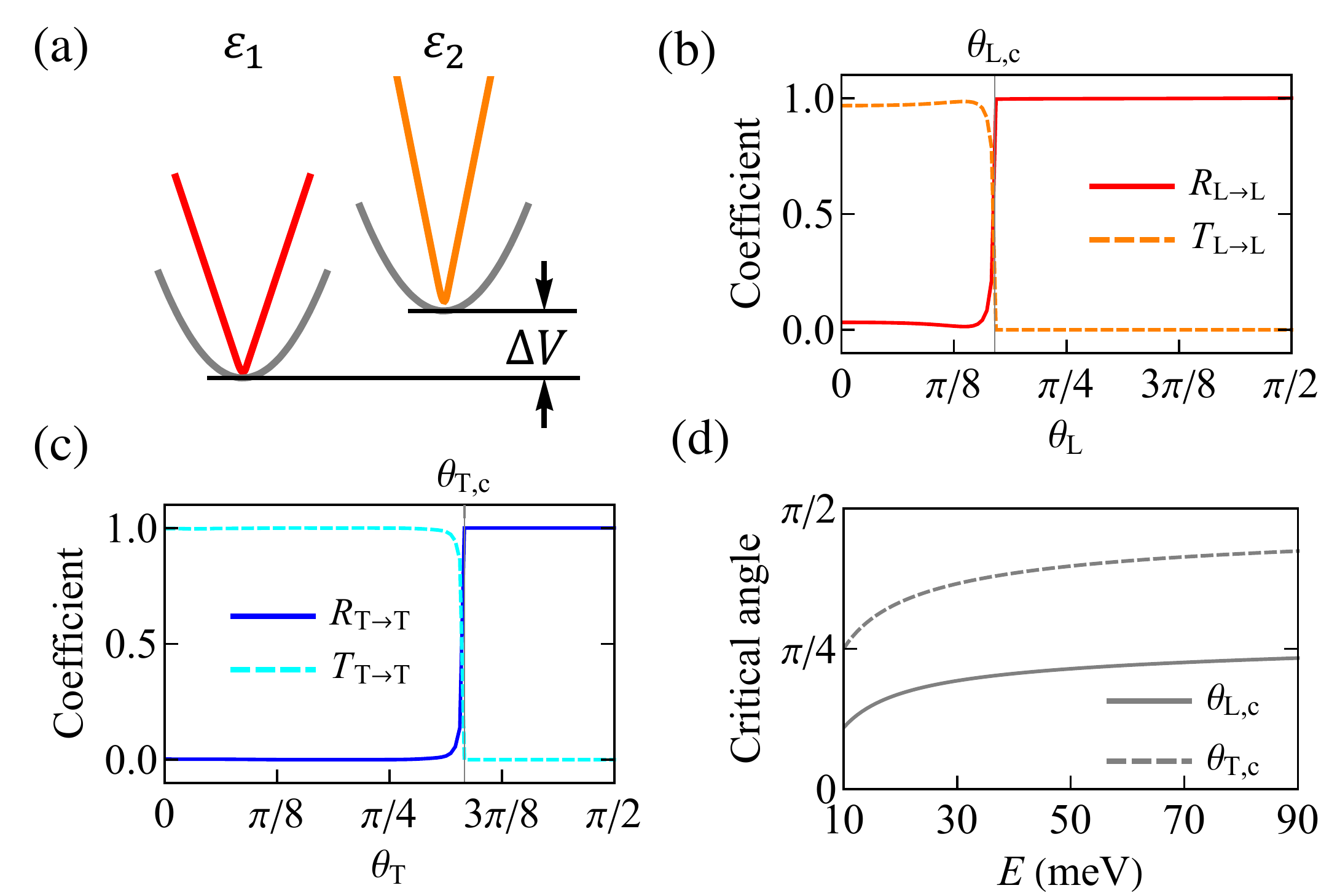}
\par\end{centering}
\caption{Effect of the energy modulation coming from the spatially varying dielectric constant. (a) Energy modulation $\Delta V_0$. For (b) L branch incidence or (c) T branch incidence, the transmission and reflection as a function of incident angle. The total internal reflection is occurred at critical angle $\theta_\text{L,c}/\theta_\text{T,c}$, respectively. (d) The critical angle as a function of the exciton energy $E$. Parameters: $\Delta V$ with amplitude of 5 meV, $J=1$ eV, $k_\text{s}=0$, and $\xi_\varepsilon=1.5$.
\label{fig:4modulation}}
\end{figure}

In the above analysis, we have assumed that the exciton energy keeps constant when modulating the dielectric environment, as the resultant changes to the exciton binding energy and electronic bandgap almost cancel each other for the 1s exciton \cite{Qiu13,Cho18,Gao16}. In realistic systems this cancellation is not exact, and there can be a modest modulation $\Delta V$ in the energy of both exciton branches, as shown in Fig.\ref{fig:4modulation} (a). We examine the effect of this energy modulation on the scattering result at a lateral interface (in Sec.IV) in the unscreened limit $k_\text{s}=0$. We adopt that $\Delta V$ has a modulation amplitude of 5 meV. Fig.\ref{fig:4modulation} (c) shows $R_{\text{T}\rightarrow\text{T}}$ and $T_{\text{T}\rightarrow\text{T}}$ as functions of the incident angle $\theta_\text{T}$. The results show that the T branch exciton is now affected by the interface as well. $T_{\text{T}\rightarrow\text{T}}=1$ for incident angles around the normal incidence. When $\theta_\text{T}$ is above a critical angle $\theta_\text{T,c}$, the total internal reflection occurs, $R_{\text{T}\rightarrow\text{T}}=1$ and $T_{\text{T}\rightarrow\text{T}}=0$. The L branch still exists the total internal reflection at the interface (in Fig.\ref{fig:4modulation} (b)) with the critical angle $\theta_\text{L,c}$ depending on $\Delta V$ and the exciton energy $E$. Fig.\ref{fig:4modulation} (d) shows the critical angles $\theta_\text{L,c}$ and $\theta_\text{T,c}$ as a function of $E$.

In the case of excitons on the 1D dielectric superlattice (in Sec.V), these spatial features of the L branch exciton's wavefunctions are not sensitive to low energy modulation (i.e. amplitude of $\Delta V$ less than 4 meV) coming from the spatially varying dielectric constant. (see Appendix~\ref{appxD} for detail)

\section{Summary}

In summary, we have put forward a design of excitonic waveguide for the longitudinal branch by placing a homogeneous monolayer TMDs on a patterned dielectric substrate, which does not affect the massive transverse branch. In a lateral dielectric interface, the L-branch exciton's transmission and reflection satisfy the Snell-Descartes law of the optical system. Nevertheless, this dielectric heterostructure appears transparent to the transverse branch. Placing the monolayer on a 1D periodic dielectric superlattice, we find that the longitudinal branch excitons have strongly renormalized dispersion and 1D spatially featured wavefunctions. In contrast, the transverse branch excitons are not affected by the substrate dielectric pattern.

The work is support by the University Grants Committee/Research Grant Council of the Hong Kong SAR (AoE/P-701/20), the Croucher Senior Research Fellowship, and HKU Seed Funding for Strategic Interdisciplinary Research. H. Y. acknowledges support by the Department of Science and Technology of Guangdong Province in China (2019QN01X061).

\appendix
\setcounter{equation}{0}

\section{Model of the monolayer TMDs exciton in periodical dielectric superlattice}\label{appxA}

We consider two bright excitons $\hat{B}_{\textbf{k},+}^{\dagger}$ and $\hat{B}_{\textbf{k},-}^{\dagger}$ which correspond to the states in opposite valleys and with opposite spins,
\begin{equation}
\begin{aligned}
X_{\textbf{k},+}\left(\textbf{r}_{\text{e}},\textbf{r}_{\text{h}}\right) & =\sum_{\textbf{q}}\Phi^{*}\left(\textbf{q}\right)\psi_{\textbf{q}+\frac{m_{\text{e}}}{M_{0}}\textbf{k},\text{c},+}\left(\textbf{r}_{\text{e}}\right)\psi_{\textbf{q}-\frac{m_{\text{h}}}{M_{0}}\textbf{k},\text{v},+}^{*}\left(\textbf{r}_{\text{h}}\right)\\
X_{\textbf{k},-}\left(\textbf{r}_{\text{e}},\textbf{r}_{\text{h}}\right) & =\sum_{\textbf{q}}\Phi\left(-\textbf{q}\right)\psi_{\textbf{q}+\frac{m_{\text{e}}}{M_{0}}\textbf{k},\text{c},-}\left(\textbf{r}_{\text{e}}\right)\psi_{\textbf{q}-\frac{m_{\text{h}}}{M_{0}}\textbf{k},\text{v},-}^{*}\left(\textbf{r}_{\text{h}}\right)
\end{aligned}
\end{equation}
where the COM momentum $\textbf{k}=\left(k_x,k_y\right)=\left(k\cos\theta,\sin\theta\right)$ and the valley pseudospin $\tau=+,-$. $\psi_{\textbf{q},\text{c}/\text{v},\tau}(\textbf{r}_{\text{e}/\text{h}})$ is the conduction/valence band Bloch state at a wave vector away from $\tau\textbf{K}$ by $\textbf{q}$, and $\Phi(\textbf{q})$ is the wavefunction for the relative motion between the electron and hole. 
The intervalley Coulomb exchange couples the two different valleys: $\hat{H}_{\text{ex,inter}}=J_{+-}\left(\textbf{k},\textbf{k}^{\prime}\right)\hat{B}_{\textbf{k},+}^{\dagger}\hat{B}^{}_{\textbf{k}^{\prime},-}+\text{h.c.}$, while the intravalley Coulomb exchange couples the same valley: $\hat{H}_{\text{ex,intra}}=\sum_\tau J_{\tau\tau}\left(\textbf{k},\textbf{k}^{\prime}\right)\hat{B}_{\textbf{k},\tau}^{\dagger}\hat{B}^{}_{\textbf{k}^{\prime},\tau}$.

The intervalley matrix element is 
\begin{equation}
\begin{aligned}
&J_{+-}\left(\textbf{k},\textbf{k}^{\prime}\right)\\
=&\int\text{d}\textbf{r}_{\text{e}}\text{d}\textbf{r}_{\text{h}} \left[\sum_{\textbf{q}}\Phi\left(\textbf{q}\right)\psi_{\textbf{q}+\frac{m_{\text{e}}}{M_{0}}\textbf{k},\text{c},+}^{*}\left(\textbf{r}_{\text{e}}\right)\psi_{\textbf{q}-\frac{m_{\text{h}}}{M_{0}}\textbf{k},\text{v},+}\left(\textbf{r}_{\text{e}}\right)\right]\\
&\times V\left(\textbf{R},\textbf{r}_{\text{eh}}\right)\\
 &\times\left[\sum_{\textbf{q}^{\prime}}\Phi\left(-\textbf{q}^{\prime}\right)\psi_{\textbf{q}^{\prime}+\frac{m_{\text{e}}}{M_{0}}\textbf{k}^{\prime},\text{c},-}\left(\textbf{r}_{\text{h}}\right)\psi_{\textbf{q}^{\prime}-\frac{m_{\text{h}}}{M_{0}}\textbf{k}^{\prime},\text{v},-}^{*}\left(\textbf{r}_{\text{h}}\right)\right]
\end{aligned}
\end{equation}
Here $V\left(\textbf{R},\textbf{r}_{\text{eh}}\right)$ is the Coulomb potential, which depends on both the relative coordinate $\textbf{r}_{\text{eh}}$ as well as the center-of-mass coordinate $\textbf{R}$ due to the spatially varying dielectric constant $\varepsilon\left(\textbf{R}\right)$. The Bloch wave functions
\begin{equation}
\begin{aligned}
\psi_{\textbf{q},\text{c}\left(\text{v}\right),+}\left(\textbf{r}_{\text{e}}\right) & =\text{e}^{{\text{i}}\left(\textbf{K}+\textbf{q}\right)\cdot\textbf{r}_{\text{e}}}u_{\textbf{q},\text{c}\left(\text{v}\right),+}\left(\textbf{r}_{\text{e}}\right)\\
&=\text{e}^{{\text{i}}\left(\textbf{K}+\textbf{q}\right)\cdot\textbf{r}_{\text{e}}}\sum_{\textbf{G}}C_{\textbf{q},\text{c}\left(\text{v}\right),+}\left(\textbf{G}\right)\text{e}^{{\text{i}}\textbf{G}\cdot\textbf{r}_{\text{e}}}\\
\psi_{\textbf{q},\text{c}\left(\text{v}\right),-}\left(\textbf{r}_{\text{h}}\right) & =\text{e}^{{\text{i}}\left(-\textbf{K}+\textbf{q}\right)\cdot\textbf{r}_{\text{h}}}u_{\textbf{q},\text{c}\left(\text{v}\right),-}\left(\textbf{r}_{\text{h}}\right)\\
&=\text{e}^{{\text{i}}\left(-\textbf{K}+\textbf{q}\right)\cdot\textbf{r}_{\text{h}}}\sum_{\textbf{G}}C_{\textbf{q},\text{c}\left(\text{v}\right),-}\left(\textbf{G}\right)\text{e}^{{\text{i}}\textbf{G}\cdot\textbf{r}_{\text{h}}}
\end{aligned}
\end{equation}
The matrix element can be written as $J_{+-}\left(\textbf{k},\textbf{k}^{\prime}\right)=\sum_{\textbf{q}\textbf{q}^{\prime}}\Phi\left(\textbf{q}\right)\Phi\left(-\textbf{q}^{\prime}\right)J_{+-}\left(\textbf{q},\textbf{q}^{\prime},\textbf{k},\textbf{k}^{\prime}\right)$ with
\begin{equation}
\begin{aligned}
 & J_{+-}\left(\textbf{q},\textbf{q}^{\prime},\textbf{k},\textbf{k}^{\prime}\right)\\
= & \int\text{d}\textbf{r}_{\text{e}}\text{d}\textbf{r}_{\text{h}}\text{e}^{{\text{i}}\left(\textbf{k}^{\prime}\cdot\textbf{r}_{\text{h}}-\textbf{k}\cdot\textbf{r}_{\text{e}}\right)}u_{\textbf{q}+\frac{m_{\text{e}}}{M_{0}}\textbf{k},\text{c},+}^{*}\left(\textbf{r}_{\text{e}}\right)u_{\textbf{q}-\frac{m_{\text{h}}}{M_{0}}\textbf{k},\text{v},+}\left(\textbf{r}_{\text{e}}\right)\\
&\times V\left(\textbf{R},\textbf{r}_{\text{e}}-\textbf{r}_{\text{h}}\right) u_{\textbf{q}^{\prime}+\frac{m_{\text{e}}}{M_{0}}\textbf{k}^{\prime},\text{c},-}\left(\textbf{r}_{\text{h}}\right)u_{\textbf{q}^{\prime}-\frac{m_{\text{h}}}{M_{0}}\textbf{k}^{\prime},\text{v},-}^{*}\left(\textbf{r}_{\text{h}}\right)\\
= & \sum_{\textbf{G},\Delta\textbf{G},\textbf{G}^{\prime},\Delta\textbf{G}^{\prime}}C_{\textbf{q}+\frac{m_{\text{e}}}{M_{0}}\textbf{k},\text{c},+}^{*}\left(\textbf{G}\right)C_{\textbf{q}-\frac{m_{\text{h}}}{M_{0}}\textbf{k},\text{v},+}\left(\textbf{G}+\Delta\textbf{G}\right)\\
&\times C_{\textbf{q}^{\prime}+\frac{m_{\text{e}}}{M_{0}}\textbf{k}^{\prime},\text{c},-}\left(\textbf{G}^{\prime}\right)C_{\textbf{q}^{\prime}-\frac{m_{\text{h}}}{M_{0}}\textbf{k}^{\prime},\text{v},-}^{*}\left(\textbf{G}^{\prime}+\Delta\textbf{G}^{\prime}\right)\\
 &\times\int\text{d}\textbf{r}_{\text{e}}\text{d}\textbf{r}_{\text{h}}V\left(\textbf{R},\textbf{r}_{\text{eh}}\right)\text{e}^{{\text{i}}\left(\Delta\textbf{G}-\textbf{k}\right)\cdot\textbf{r}_{\text{e}}-{\text{i}}\left(\Delta\textbf{G}^{\prime}-\textbf{k}^{\prime}\right)\cdot\textbf{r}_{\text{h}}}
\end{aligned}
\end{equation}

Note that $\text{e}^{{\text{i}}\left(\Delta\textbf{G}-\textbf{k}\right)\cdot\textbf{r}_{\text{e}}-{\text{i}}\left(\Delta\textbf{G}^{\prime}-\textbf{k}^{\prime}\right)\cdot\textbf{r}_{\text{h}}}=\text{e}^{{\text{i}}\left[\frac{m_{\text{h}}}{M_{0}}\left(\Delta\textbf{G}-\textbf{k}\right)+\frac{m_{\text{e}}}{M_{0}}\left(\Delta\textbf{G}^{\prime}-\textbf{k}^{\prime}\right)\right]\cdot\textbf{r}_{\text{eh}}}\text{e}^{{\text{i}}\left(\Delta\textbf{G}-\textbf{k}-\Delta\textbf{G}^{\prime}+\textbf{k}^{\prime}\right)\cdot\textbf{R}}$. Since $V\left(\textbf{R},\textbf{r}_{\text{eh}}\right)$ varies very slowly with $\textbf{R}$, so $\int\text{d}\textbf{R}V\left(\textbf{R},\textbf{r}_{\text{eh}}\right)\text{e}^{{\text{i}}\left(\Delta\textbf{G}-\textbf{k}-\Delta\textbf{G}^{\prime}+\textbf{k}^{\prime}\right)\cdot\textbf{R}}=\delta_{\Delta\textbf{G},\Delta\textbf{G}^{\prime}}\int\text{d}\textbf{R}V\left(\textbf{R},\textbf{r}_{\text{eh}}\right)\text{e}^{-{\text{i}}\left(\textbf{k}-\textbf{k}^{\prime}\right)\cdot\textbf{R}}$.
So
\begin{equation}
\begin{aligned}
&J_{+-}\left(\textbf{q},\textbf{q}^{\prime},\textbf{k},\textbf{k}^{\prime}\right)\\
= & \sum_{\textbf{G},\textbf{G}^{\prime},\Delta\textbf{G}}C_{\textbf{q}+\frac{m_{\text{e}}}{M_{0}}\textbf{k},\text{c},+}^{*}\left(\textbf{G}\right)C_{\textbf{q}-\frac{m_{\text{h}}}{M_{0}}\textbf{k},\text{v},+}\left(\textbf{G}+\Delta\textbf{G}\right)\\
&\times C_{\textbf{q}^{\prime}+\frac{m_{\text{e}}}{M_{0}}\textbf{k}^{\prime},\text{c},-}\left(\textbf{G}^{\prime}\right)C_{\textbf{q}^{\prime}-\frac{m_{\text{h}}}{M_{0}}\textbf{k}^{\prime},\text{v},-}^{*}\left(\textbf{G}^{\prime}+\Delta\textbf{G}\right)\\
 &\times\int\text{d}\textbf{R}\text{d}\textbf{r}_{\text{eh}}V\left(\textbf{R},\textbf{r}_{\text{eh}}\right)\text{e}^{-{\text{i}}\left(\textbf{k}-\textbf{k}^{\prime}\right)\cdot\textbf{R}}\text{e}^{{\text{i}}\left(\Delta\textbf{G}-\frac{m_{\text{h}}}{M_{0}}\textbf{k}-\frac{m_{\text{e}}}{M_{0}}\textbf{k}^{\prime}\right)\cdot\textbf{r}_{\text{eh}}}\\
= & J_{+-}^{\text{LR}}\left(\textbf{q},\textbf{q}^{\prime},\textbf{k},\textbf{k}^{\prime}\right)+J_{+-}^{\text{SR}}\left(\textbf{q},\textbf{q}^{\prime},\textbf{k},\textbf{k}^{\prime}\right)
\end{aligned}
\end{equation}
Here the short-range part $J_{+-}^{\text{SR}}$ can be ignored as it is much smaller than $J_{+-}^{\text{LR}}$. The long-range part is
\begin{equation}
\begin{aligned}
 J^\text{LR}_{+-}\left(\textbf{q},\textbf{q}^{\prime},\textbf{k},\textbf{k}^{\prime}\right)=&\frac{1}{A}\left\langle u_{\textbf{q}+\frac{m_{\text{e}}}{M_{0}}\textbf{k},\text{c},+}\right|\left.u_{\textbf{q}-\frac{m_{\text{h}}}{M_{0}}\textbf{k},\text{v},+}\right\rangle\\
 &\times\left\langle u_{\textbf{q}^{\prime}-\frac{m_{\text{h}}}{M_{0}}\textbf{k}^{\prime},\text{v},-}\right|\left.u_{\textbf{q}^{\prime}+\frac{m_{\text{e}}}{M_{0}}\textbf{k}^{\prime},\text{c},-}\right\rangle\\
&\times V(\textbf{k}-\textbf{k}^\prime,\frac{m_\text{h}}{M_0}\textbf{k}+\frac{m_\text{e}}{M_0}\textbf{k}^\prime)
\end{aligned}
\end{equation}
and $V(\textbf{Q},\textbf{q})\equiv\int\text{d}\textbf{r}_{\text{eh}}\text{d}\textbf{R}\text{e}^{-{\text{i}}\textbf{q}\cdot\textbf{r}_{\text{eh}}}\text{e}^{-{\text{i}}\textbf{Q}\cdot\textbf{R}}V\left(\textbf{R},\textbf{r}_{\text{eh}}\right)$. We use the $\textbf{k}\cdot\textbf{p}$ expansion to evaluate the matrix element (and $m_\text{e}=m_\text{h}=m$),
\begin{equation}
\begin{aligned}
\left|u_{\textbf{q}+\frac{\textbf{k}}{2},\text{c},\tau}\right\rangle=&\left|u_{\textbf{q},\text{c},\tau}\right\rangle+\frac{\hslash}{2m}\textbf{k}\cdot\sum_{\text{n}\neq\text{c}}\frac{\left\langle u_{\textbf{q},\text{n},\tau}\right|\hat{\textbf{p}}\left|u_{\textbf{q},\text{c},\tau}\right\rangle}{E_{\textbf{q},\text{c},\tau}-E_{\textbf{q},\text{n},\tau}}\left|u_{\textbf{q},\text{n},\tau}\right\rangle\\
&+\mathcal{O}(k^2)\\
\left|u_{\textbf{q}-\frac{\textbf{k}}{2},\text{v},\tau}\right\rangle=&\left|u_{\textbf{q},\text{v},\tau}\right\rangle+\frac{\hslash}{2m}\textbf{k}\cdot\sum_{\text{n}\neq\text{v}}\frac{\left\langle u_{\textbf{q},\text{n},\tau}\right|\hat{\textbf{p}}\left|u_{\textbf{q},\text{v},\tau}\right\rangle}{E_{\textbf{q},\text{n},\tau}-E_{\textbf{q},\text{v},\tau}}\left|u_{\textbf{q},\text{n},\tau}\right\rangle\\
&+\mathcal{O}(k^2)
\end{aligned}
\end{equation}
Then $\left\langle u_{\textbf{q}+\frac{\textbf{k}}{2},\text{c},\tau}\left|u_{\textbf{q}-\frac{\textbf{k}}{2},\text{v},\tau}\right\rangle\right.=\frac{\hslash}{m}\textbf{k}\cdot\frac{\left\langle u_{\textbf{q},\text{c},\tau}\right|\hat{\textbf{p}}\left|u_{\textbf{q},\text{v},\tau}\right\rangle}{E_{\textbf{q},\text{c},\tau}-E_{\textbf{q},\text{v},\tau}}+\mathcal{O}(k^2)$. We focus on the bright exciton within and near the light cone, thus we only need to keep the leading term in $\textbf{k}$.  The optical dipole matrix element is $\textbf{d}_{\text{cv},\textbf{q},\tau}\equiv\frac{\text{i}\hslash}{m}\frac{\left\langle u_{\textbf{q},\text{c},\tau}\right|\hat{\textbf{p}}\left|u_{\textbf{q},\text{v},\tau}\right\rangle}{E_{\textbf{q},\text{c},\tau}-E_{\textbf{q},\text{v},\tau}}$.
So the intervalley matrix element is
\begin{equation}
J_{+-}\left(\textbf{k},\textbf{k}^{\prime}\right)=\left(\textbf{k}\cdot\textbf{d}^{}_{\text{X},+}\right)\left(\textbf{k}^{\prime}\cdot\textbf{d}^*_{\text{X},-}\right)V(\textbf{k}-\textbf{k}^\prime,\frac{m_\text{h}}{M_0}\textbf{k}+\frac{m_\text{e}}{M_0}\textbf{k}^\prime)
\end{equation}
where the exciton transition dipole is $\textbf{d}_{\text{X},+}=\textbf{d}_{\text{X},-}^{*}=\frac{1}{\sqrt{A}}\sum_{\textbf{q}}\Phi\left(\textbf{q}\right)\textbf{d}_{\text{cv},\textbf{q},+}$. Similarly we can get the intervalley and intravalley matrix element as
\begin{equation}
    J_{\tau,\tau^\prime}\left(\textbf{k},\textbf{k}^{\prime}\right)=\left(\textbf{k}\cdot\textbf{d}^{}_{\text{X},\tau}\right)\left(\textbf{k}^{\prime}\cdot\textbf{d}^*_{\text{X},\tau^\prime}\right)V(\textbf{k}-\textbf{k}^\prime,\frac{m_\text{h}}{M_0}\textbf{k}+\frac{m_\text{e}}{M_0}\textbf{k}^\prime)
\end{equation}

\section{Patterned dielectric superlattice}\label{appxB}

\begin{figure}[h]
    \begin{centering}
    \includegraphics[scale=0.28]{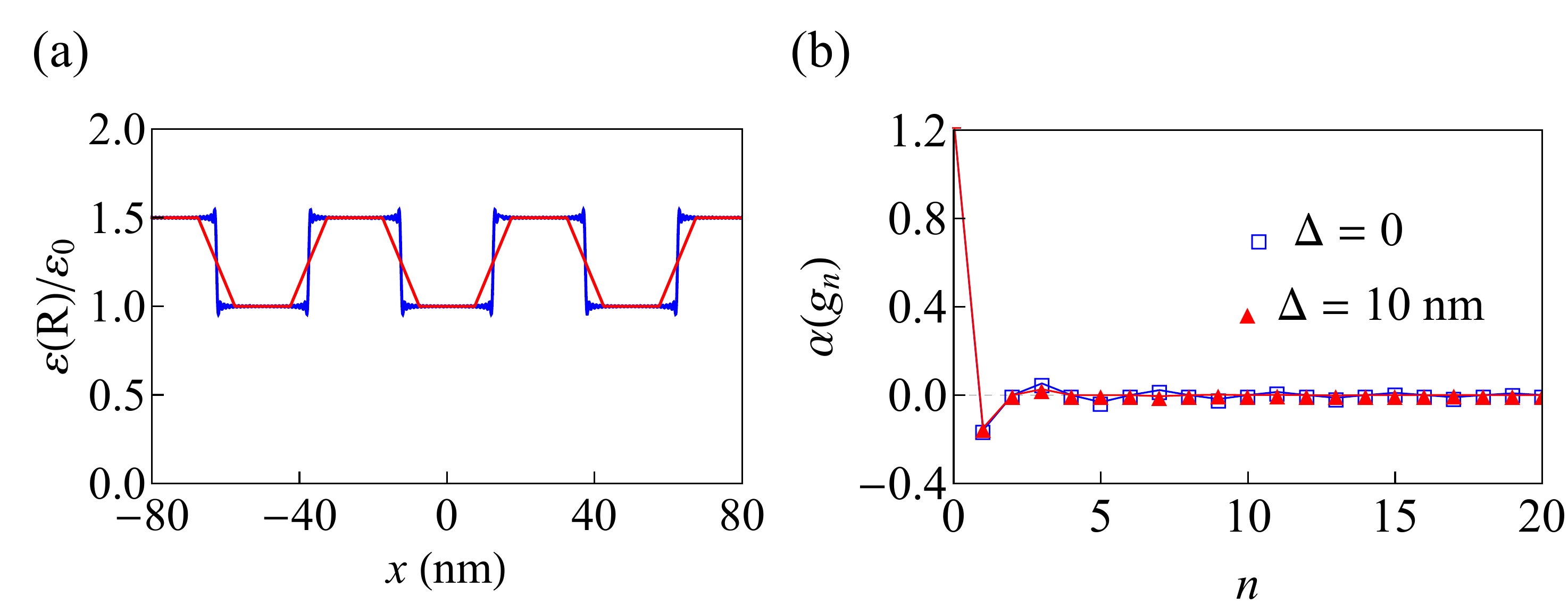}
    \par\end{centering}
    \caption{(a) Numerical result of periodic dielectric $\varepsilon\left(\textbf{R}\right)/\varepsilon_{0}$ as a function of $x-$axis. Parameter: $\xi=1.5$, $\Delta=0$ (blue line) and $\Delta=10$ nm (red line). (b) The corresponding coefficients in Fourier transform. 
    \label{fig-3-5}}
\end{figure}

Considering the local dielectric constant $\varepsilon(\textbf{R})$ (in Eq.(\ref{eq:1Dep})) changes smoothly at the lateral interfaces in a length scale $\Delta=10$ nm. We transfer the spatial distribution $\varepsilon(\textbf{R})$ to momentum space using Fourier transform $\varepsilon\left(\textbf{R}\right)^{-1}=\varepsilon_{0}^{-1}\sum_{\bm{g}_{n}}\alpha\left(\textbf{g}_{n}\right){\rm e}^{{\rm i}\textbf{g}_{n}\cdot\textbf{r}}$. Here the wave vector $\textbf{g}_{n}=\left(\frac{2\pi}{a_1+a_2}n,0\right)$ and corresponding coefficient
\begin{equation}
\alpha\left(\textbf{g}_{n}\right)=\begin{cases}
\begin{array}{ccc}
\frac{\xi_\varepsilon a_1+a_2}{\xi_\varepsilon\left(a_1+a_2\right)}, &  & n=0\\
\frac{2(1-\xi_\varepsilon)}{\xi_\varepsilon \Delta\left(a_1+a_2\right)}\frac{\cos\left(\frac{a_1+\Delta}{2}g_{n}\right)-\cos\left(\frac{a_1-\Delta}{2}g_{n}\right)}{g_{n}^{2}}, &  & n\neq0
\end{array} & .\end{cases}\label{eq:1DFourierC-1}
\end{equation}
The quasi-1D pattern dielectric superlattice at $\xi_\varepsilon=1.5$ with the smooth lateral interface $\Delta$ are shown in Fig.\ref{fig-3-5} (a), while (b) presents the corresponding coefficients.

\section{Scattering analysis in a lateral dielectric interface}\label{appxC}

At a lateral dielectric interface, the Hamiltonian of a exciton at each region is Eq.(10), where the eigenvalue is $E_\text{L}\left(\bm{k}\right)=\frac{\hbar^{2}}{2M_{0}}k^2+\frac{2J_\eta}{K}\frac{k^2}{k+k_\text{s}}$ or $E_\text{T}\left(\bm{k}\right)=\frac{\hbar^{2}}{2M_{0}}k^2$ with eigenvector $\Psi_\text{L/T}\left(\bm{r}\right)=\frac{1}{\sqrt{2}}\text{e}^{{\rm i}\bm{k}\cdot\bm{r}}\left\{ 1,\mp\text{e}^{2{\rm i}\theta}\right\} ^{\text{T}}$, and $J_\eta=\frac{\varepsilon_0}{2\varepsilon_\eta}J$. For the whole lateral dielectric interface, the valley-orbit coupling is coordinate dependent $J(x)$ due to the form of dielectric Eq.(\ref{eq:1Dep}). $J\left(x\right)$ equals to two different constant $J_{2}$ and $J_{1}$ at two regions of the device, respectively. Firstly, we consider the width of the transition region $\Delta=0$ and $k_\text{s}=0$. We assume the L branch incidence with incident angle $\theta_\text{\text{L}}$, the exciton is regarded as a quasi-particle with wavefunction 
\begin{widetext}
\begin{equation}
\Psi\left(\bm{r}\right)=\begin{cases}
\begin{array}{lc}
\frac{1}{\sqrt{2}}{\rm e}^{{\rm i}k_{y}y}
\left[{\rm e}^{{\rm i}k_{x,\text{L}}x}
\left(\begin{array}{c}1\\-{\rm e}^{2{\rm i}\theta_{\text{L}}}\end{array}\right)
+r_{\text{L}}{\rm e}^{-{\rm i}k_{x,\text{L}}x}
\left(\begin{array}{c}1\\-{\rm e}^{2{\rm i}\left(\pi-\theta_{\text{L}}\right)}\end{array}\right)+r_{\text{T}}{\rm e}^{-{\rm i}k_{x,\text{T}}x}
\left(\begin{array}{c}1\\{\rm e}^{2{\rm i}\left(\pi-\theta_{\text{T}}\right)}\end{array}\right)
\right], & x<0\\
\frac{1}{\sqrt{2}}{\rm e}^{{\rm i}k_{y}y}\left[t_{\text{L}}{\rm e}^{{\rm i}q_{x,\text{L}}x}\left(\begin{array}{c}
1\\
-{\rm e}^{2{\rm i}\phi_{\text{L}}}
\end{array}\right)+t_{\text{T}}{\rm e}^{{\rm i}q_{x,\text{T}}x}\left(\begin{array}{c}
1\\
{\rm e}^{2{\rm i}\phi_{\text{T}}}
\end{array}\right)\right], & x>0
\end{array}\end{cases}
\label{eq:tmdwf1}
\end{equation}
\end{widetext}
here the kinetic momentum of the exciton satisfies
\begin{equation}
\begin{aligned}
k_{\text{L}} & =\sqrt{\frac{4M_0^2J_2^2}{\hbar^4K^2}+\frac{2EM_0}{\hbar^2}}-\frac{2J_2M_0}{\hbar^2M_0}, \\
q_{\text{L}} & =\sqrt{\frac{4M_0^2J_1^2}{\hbar^4K^2}+\frac{2EM_0}{\hbar^2}}-\frac{2J_1M_0}{\hbar^2M_0},\\
k_{\text{T}}&=q_{\text{T}}=\sqrt{\frac{2M_{0}E}{\hslash^{2}}}, \quad k_{y}=k_{\text{L}}\sin\theta_{\text{L}}\\
k_{x,\text{L}}&=k_{\text{L}}\cos\theta_{\text{L}}, \quad k_{x,\text{T}}=k_{\text{T}}\cos\theta_{\text{T}}, \quad q_{x,\text{L}}=q_{\text{L}}\cos\phi_{\text{L}}\\
\theta_{\text{T}}&=\arcsin\left(\frac{k_{\text{L}}}{k_{\text{T}}}\sin\theta_{\text{L}}\right), \quad \phi_{\text{L}}=\arcsin\left(\frac{k_{\text{L}}}{q_{\text{L}}}\sin\theta_{\text{L}}\right)\\
\phi_{\text{T}}&=\arcsin\left(\frac{k_{\text{L}}}{q_{\text{T}}}\sin\theta_{\text{L}}\right)\\
\label{eq:tmdsharpk1}
\end{aligned}
\end{equation}
The boundary conditions at the interface point $x_{0}$ (i.e. $x_0=0$),
\begin{equation}
\begin{aligned}
\left.\hat{\beta}\left(J_2\right)\sqrt{\frac{J_2}{k_{x,\text{L}}}}\Psi\left(\bm{r}\right)\right|_{x=x_{0}^{-}} & =\left.\hat{\beta}\left(J_1\right)\sqrt{\frac{J_1}{q_{x,\text{L}}}}\Psi\left(\bm{r}\right)\right|_{x=x_{0}^{+}}\\
\sqrt{\frac{J_2}{k_{x,\text{L}}}}\Psi\left(x_{0}^{-}\right) & =\sqrt{\frac{J_1}{q_{x,\text{L}}}}\Psi\left(x_{0}^{+}\right)
\label{eq:tmdbc1}
\end{aligned}
\end{equation}
where operator $\hat{\beta}$ in real space has
\begin{equation}
\begin{aligned}
\hat{\beta}\left(J_\eta\right)=\left(\begin{array}{cc}
-\left(\frac{\hslash^{2}}{2m}\frac{k_{\eta,x}}{J_\eta}+\frac{1}{K}\right)\frac{\partial}{\partial x} & \frac{1}{K}\left(\frac{\partial}{\partial x}+2k_{y}\right)\\
\frac{1}{K}\left(\frac{\partial}{\partial x}-2k_{y}\right) & -\left(\frac{\hslash^{2}}{2m}\frac{k_{\eta,x}}{J_\eta}+\frac{1}{K}\right)\frac{\partial}{\partial x}
\end{array}\right).
\end{aligned}
\end{equation}
Combining the boundary condition Eq.(\ref{eq:tmdbc1}) with wavefunction Eq.(\ref{eq:tmdwf1}), the amplitude $\left\{ r_{\text{L}},r_{\text{T}},t_{\text{L}},t_{\text{T}}\right\}$ and reflection/ transmission coefficient $R_{\chi\rightarrow\chi^\prime}/T_{\chi\rightarrow\chi^\prime}$ are calcualted.
\begin{figure}[h]
\begin{centering}
\includegraphics[scale=0.32]{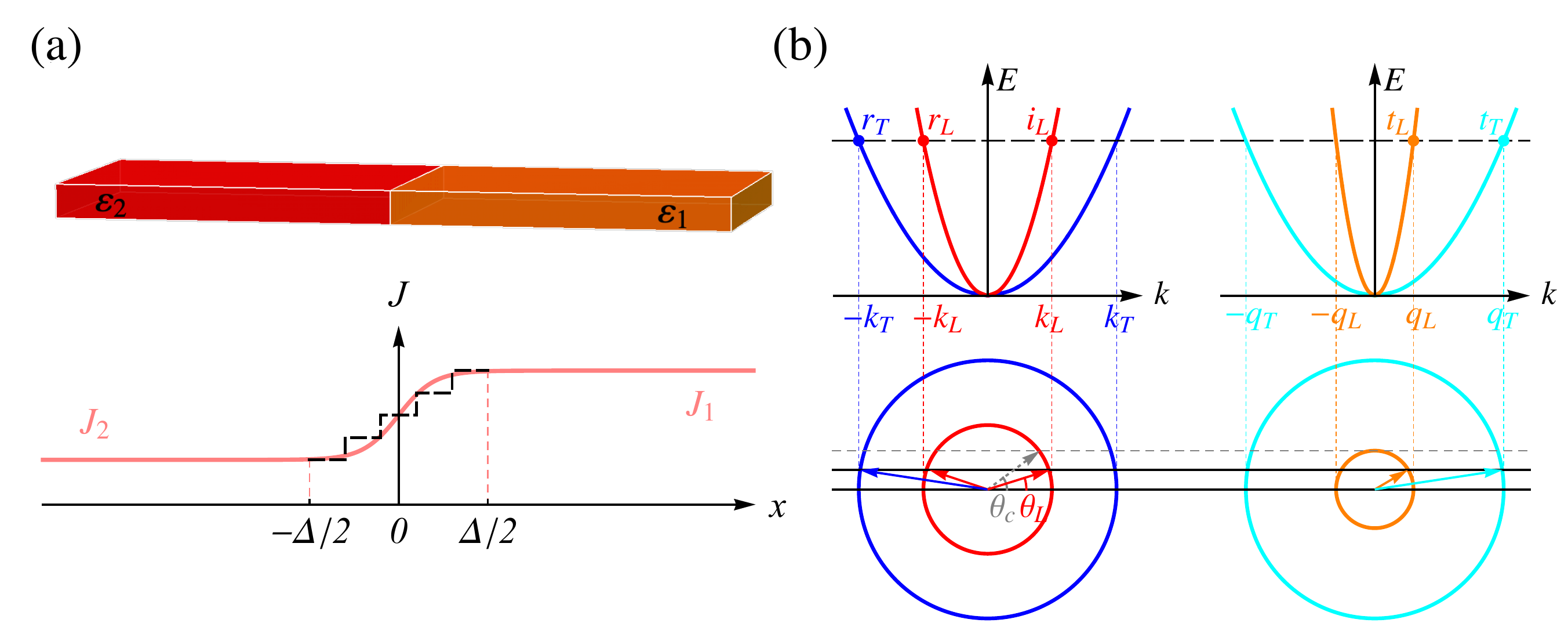}
\par\end{centering}
\caption{(a) Scheme of 1D patterned superlattice substrate, which provides a spatially modulating strength of valley-orbit coupling $J\left(x\right)$. $J\left(x\right)$ approaches to two different constant $J_{2}$ and $J_{1}$ at two regions, respectively. Here is a $\Delta$-width transitional region at the interface. (b) Schematic energy dispersion of the exciton at a lateral dielectric interface. Transmission and reflection coefficients at the L branch incidence is denoted.  \label{fig:j1}}
\end{figure}

\begin{figure}
\begin{centering}
\includegraphics[scale=0.37]{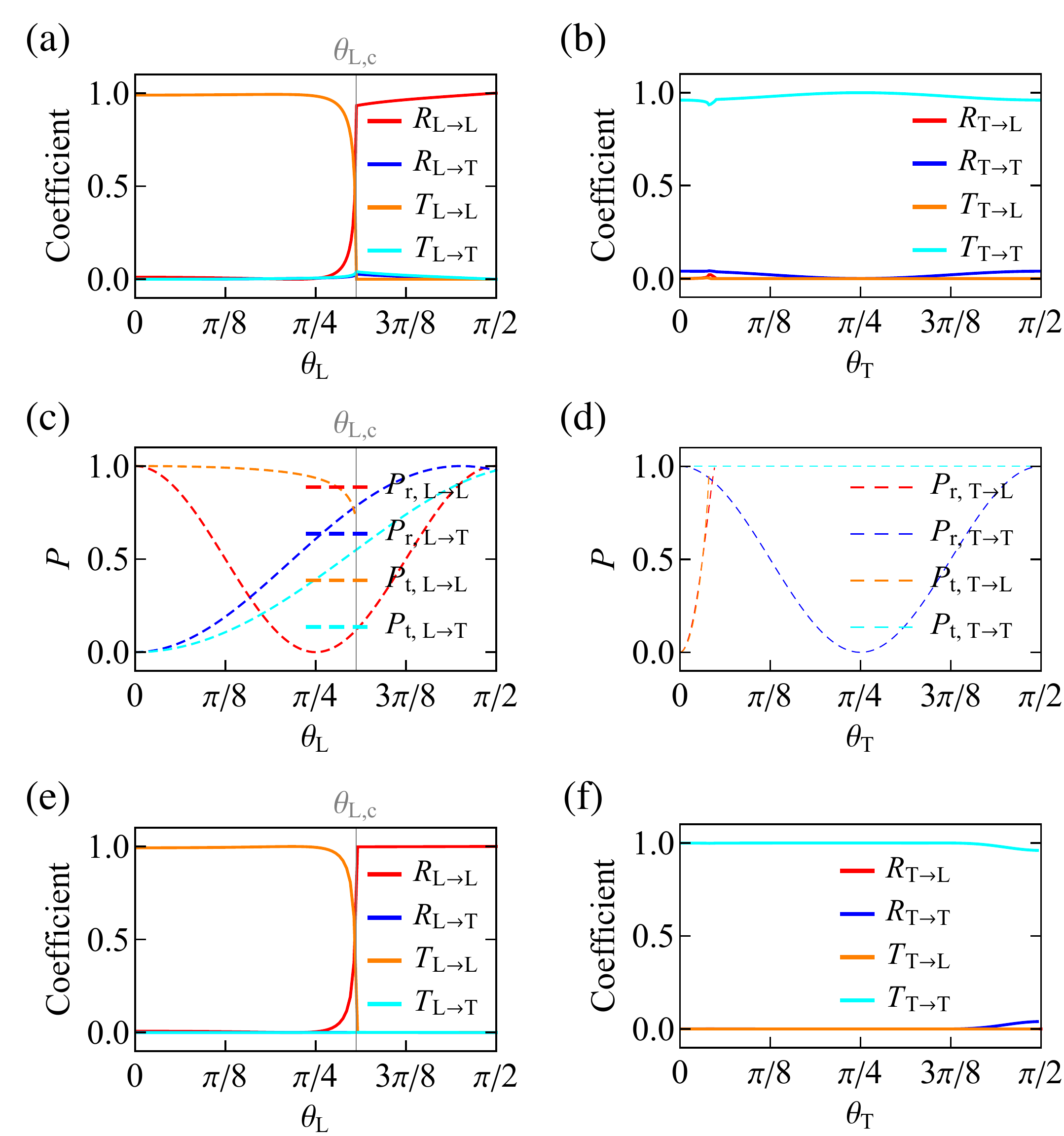}
\par\end{centering}
\caption{Numerical results of scattering at a lateral interface where the two adjacent segments have distinct dielectric constants. Here, the L branch transmission/reflection coefficients are denoted by $T_{\chi\rightarrow\text{\text{L}}}$ (orange solid line)/$R_{\chi\rightarrow\text{\text{L}}}$ (red solid line), while the T branch transmission coefficient $T_{\chi\rightarrow\text{\text{T}}}$ (cyan solid line) and reflection $R_{\chi\rightarrow\text{\text{T}}}$ (blue solid line). The branch index of incident exciton is $\chi=\text{\text{L}}/\text{\text{T}}$. The pseudospin projection is marked by corresponding dashed line, respectively. In the sharp interface case, (a) the transmission, reflection coefficient and (c) pseudospin projection $P_{\mu,\sigma}$ as functions of the incident angle $\theta_\text{\text{L}}$ at the L branch incidence. (b,d) The T branch incidence with incident angle $\theta_\text{\text{T}}$. (e,f) The scattering result with nonzero transition region at interface. Parameters: $E=5$ meV, $J=1$ eV, $k_\text{s}=0$, $\xi=1.5$, and transitional region $\Delta=20$ nm.
\label{fig:scat}}
\end{figure}

Based on the conservation of spin in scattering process, these numerical results of both the L and T branch incidence are understood through the pseudospin texture analysis. The projection of reflected/transmitted wave's pseudospin to incident wave's pseudospin $P_{\text{r}/\text{\text{t}},\chi\rightarrow\chi^\prime}=\left|\left\langle\chi_i\left|\chi^\prime_{\text{r}/\text{\text{t}}}\right\rangle\right.\right|^2$. $\chi_\mu$ is the valley pseudospin part of exciton wavefunction, and $\chi_\mu$. $\mu\in\{\text{i},\text{r},\text{t}\}$ stands for the incident, reflected or transmitted wave, respectively. Fig.\ref{fig:scat} (c, d) show the pseudospin projection as the function of incident angle. At the normal incidence of L branch exciton $\theta_{\text{L}}=0$ in (c), the pseudospins of the incident exciton and the intra-branch scattering exciton point in the same direction, i.e. $P_{\text{r},\text{\text{L}}\rightarrow\text{\text{L}}}=P_{\text{\text{t}},\text{\text{L}}\rightarrow\text{\text{L}}}=1$. But they point in opposite directions for the inter-branch scattering ($P_{\text{r},\text{\text{L}}\rightarrow\text{\text{T}}}=P_{\text{\text{t}},\text{\text{L}}\rightarrow\text{\text{T}}}=0$).  It leads to that the intra-branch scattering occupies the total output, while the inter-branch scattering is quite weak. 
When $\theta_\text{\text{L}}$ beyonds the critical angle $\theta_\text{L,c}$, the oscillating transmitted wave of the L branch vanishes, and total internal reflection emerges. The numerical result of the T branch incidence is shown in Fig.\ref{fig:scat} (d). When the incident angle $\theta_\text{T}$ changes from 0 to $\pi/2$, the pseudospin of the incident exciton and the T branch transmitted exciton always point to the same direction, i.e. $P_{\text{\text{t}},\text{\text{T}}\rightarrow\text{\text{T}}}=1$. It indicates that the intra-branch transmission coefficient $T_{\text{\text{T}}\rightarrow\text{\text{T}}}$ plays the dominate role of outputs.

We consider a $\Delta$-width transitional region where the local dielectric constant changes continuously from the value of the left region to that value of the right region. In the scattering calculation, this transitional region is divided into $n$ equal pieces of small step functions, at which the wavefunctions are matched with the boundary conditions. The excitons feel the dielectric constant $\varepsilon_i=\left[i\varepsilon_1+(n-i)\varepsilon_2\right]/n$ at the $i$th piece. The valley-orbital coupling amplitude has $J_i=\frac{\varepsilon_0}{2\varepsilon_\eta}J$. For each region, the two-component wavefunction of exciton $\Psi_{i}\left(\bm{r}\right)=\left\{ \psi_{1,i}\left(\bm{r}\right),\psi_{2,i}\left(\bm{r}\right)\right\} ^{\text{T}}$,
\begin{equation}
\begin{aligned}
\psi_{1,i}\left(\bm{r}\right)=\frac{1}{\sqrt{2}}\text{e}^{{\rm i}k_{y}y}&\left(a_{i}\text{e}^{{\rm i}k_{x,\text{L},i}x}+b_{i}\text{e}^{-{\rm i}k_{x,\text{L},i}x}\right.\\
&\quad\left.+c_{i}\text{e}^{{\rm i}k_{x,\text{T}}x}+d_{i}\text{e}^{-{\rm i}k_{x,\text{T}}x}\right)\\
\psi_{2,i}\left(\bm{r}\right)=\frac{1}{\sqrt{2}}\text{e}^{{\rm i}k_{y}y}&\left(-a_{i}\text{e}^{{\rm i}k_{x,\text{L},i}x+2{\rm i}\theta_{\text{L},i}}-b_{i}\text{e}^{-{\rm i}k_{x,\text{L},i}x-2{\rm i}\theta_{\text{L},i}}\right.\\
&\quad\left.+c_{i}\text{e}^{{\rm i}k_{x,\text{T}}x+2{\rm i}\theta_{\text{T}}}+d_{i}\text{e}^{-{\rm i}k_{x,\text{T}}x-2{\rm i}\theta_{\text{T}}}\right)
\label{eq:tmdwf2}
\end{aligned}
\end{equation}
where the kinetic momentum
\begin{equation}
\begin{aligned}
k_{\text{L},i} & =\sqrt{\frac{4M_0^2J_i^2}{\hbar^4K^2}+\frac{2EM_0}{\hbar^2}}-\frac{2J_iM_0}{\hbar^2M_0}, k_{\text{T}} =\sqrt{\frac{2M_{0}E}{\hslash^{2}}},\\
k_{x,\text{L},i}&=k_{\text{L},i}\cos\theta_{\text{L},i},\quad\theta_{\text{L},i}=\arcsin\left(\frac{k_{\text{L},0}}{k_{\text{L},i}}\sin\theta_{\text{L},0}\right),\\
k_{x,\text{T}}&=k_{\text{T}}\cos\theta_{\text{T}},\quad\theta_{\text{\text{T}}}=\arcsin\left(\frac{k_{\text{L},0}}{k_{\text{T}}}\sin\theta_{\text{L},0}\right)
\label{eq:kvalue}
\end{aligned}
\end{equation}
where incident angle $\theta_{\text{L},0}=\theta_{\text{L}}$, $a_{0}=1$, and $b_{n}=d_{n}=0$. The transmission and reflection coefficients are determined via the boundary conditions at all the joint points. The numerical scattering result of the L branch incident with the incident angle $\theta_\text{\text{L}}$ is shown in Fig.\ref{fig:scat} (e). The dominant role of intra-branch scattering output ($R_{\text{\text{L}}\rightarrow\text{\text{L}}}+T_{\text{\text{L}}\rightarrow\text{\text{L}}}$) is enhanced. At the T branch incidence in (f), $T_\text{\text{T}}$ approaches to 1, while $R_\text{\text{T}}$ is suppressed evidently. It illustrates that the dielectric heterostructure seems nearly transparent for the propagating exciton of the T branch.

\section{Robustness of the 1D features of exciton's wavefunciton}\label{appxD}

\begin{figure}
\begin{centering}
\includegraphics[scale=0.45]{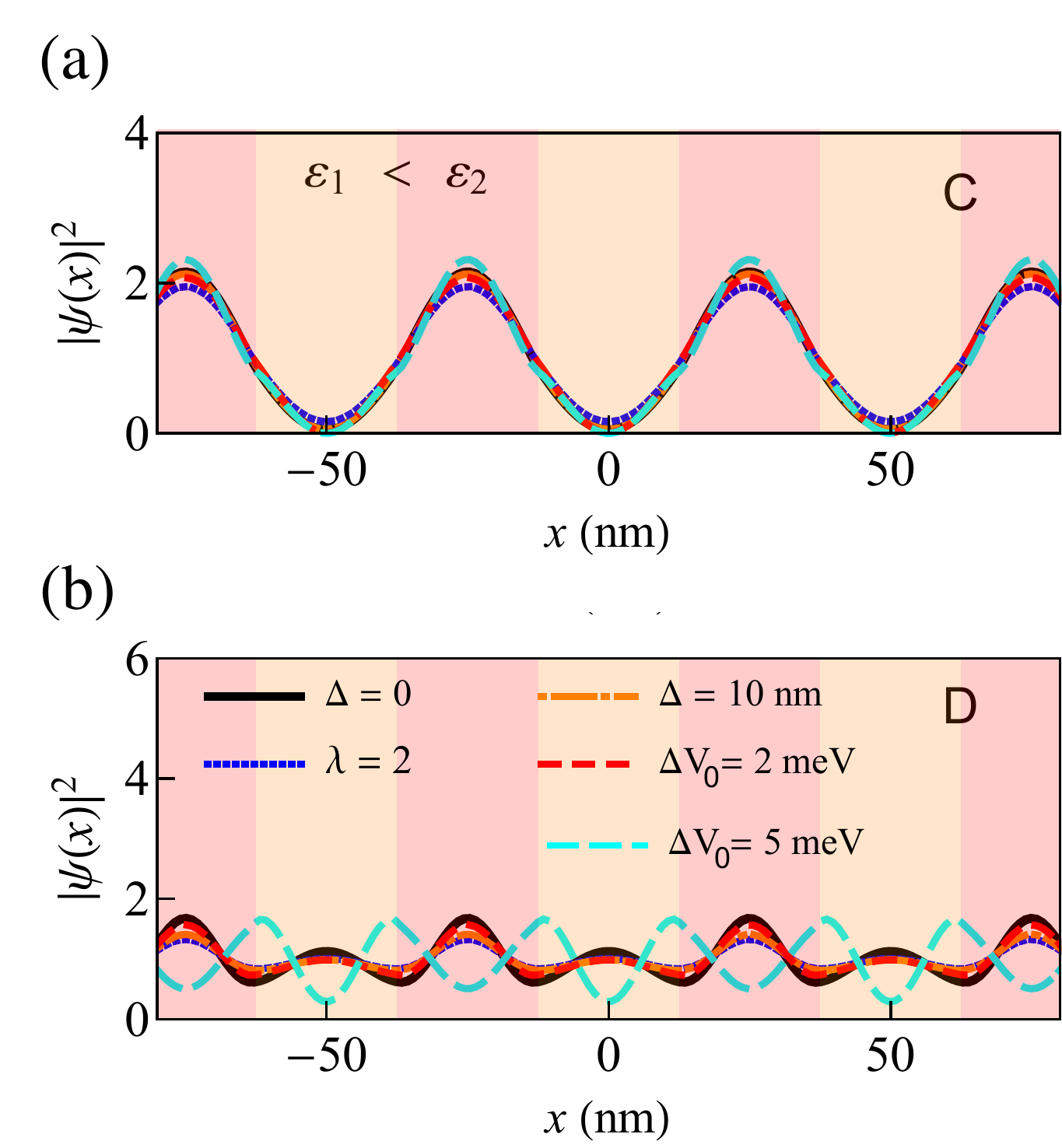}
\par\end{centering}
\caption{Probability distribution of the L branch exciton with 1D periodic dielectric superlattice at momentum points C and D (in Fig.3 (a)). The alternating colored regions denote the different dielectric constants $\varepsilon_1$ and $\varepsilon_2$. Black solid line stands for $\Delta=0$ case. Orange dotted dashed line stands for the $\Delta=10$ nm case. The blue dotted line is the screening wavevector with $\lambda=2$ case. The red (blue) dashed line is the case of energy modulation $\Delta V_0=2$ meV (5meV).
\label{fig:valid}}
\end{figure}

Based on the expression of the exchange Hamiltonian Eq.(\ref{eq:exH1}), the different form of the Coulomb interactions does not change the previous main results. Considering a dielectric superlattice $\varepsilon(\textbf{R})$ in Eq.(\ref{eq:1Dep}) with $\Delta=10$ nm and $\xi=1.5$, the exciton density of the L branch at C and D points (the momentum points marked in Fig.3 (a)) are denoted by the grey dash-dotted lines (in Fig.\ref{fig:valid}). Compared to the $\Delta=0$ case (black solid lines), it illustrates quantitatively that the main results are independent of the Coulomb interactions. Similarly, the parameter $\lambda$ of the screening wavevector $k_0(\textbf{g})=\lambda g$ in Eq.(\ref{eq:exH1}) is changed (for instance, $\lambda=2$). The  results are denoted by the blue dotted lines in Fig.\ref{fig:valid} (b), and the density distribution is almost unchanged.

Due to the cancellation of the changes to the exciton binding energy and electronic bandgap, we have ignored the exciton energy changing when modulating the dielectric environment. In realistic systems, both the $\textbf{K}$ and $-\textbf{K}$ valley excitons feel a residue small energy modulation described by $\Delta\hat{H}=\sum_{\textbf{k},\textbf{g}}\Delta V(\textbf{g})\left(\hat{B}^\dagger_{\textbf{k},+}\hat{B}^{}_{\textbf{k}+\textbf{g},+}+\hat{B}^\dagger_{\textbf{k},-}\hat{B}^{}_{\textbf{k}+\textbf{g},-}\right)$, where $\Delta V(\textbf{g})$ is the Fourier coefficient of the modulated potential in real space $\Delta V(\textbf{R})$. $\Delta V_0$ denotes the different of $\Delta V(\textbf{R})$ at different dielectric constant regions. We find that the density distribution (marked by red dashed line in Fig.\ref{fig:valid}) is almost unchanged when the potential modulated slightly $\Delta V_0=2$ meV. But at $\Delta V_0=5$ meV (cyan dashed line), such a significant potential modulation causes that the mode which concentrates in low-$\varepsilon$ regions performs a higher energy than the mode of high-$\varepsilon$ dominant.

\bibliographystyle{apsrev4-1}

\end{document}